# Optical activity of chiral excitons

Peter C. Sercel[1]*, Matthew P. Hautzinger[2], Ruyi Song[3], Volker Blum[4], Matthew C. Beard[2]

[1]Center for Hybrid Organic–Inorganic Semiconductors for Energy, Golden, Colorado 80401, USA.
[2]National Renewable Energy Laboratory, Golden, CO 80401, USA
[3]Dept. of Chemistry, Duke University, Durham, NC 27708, USA
[4]Dept. of Mechanical Engineering and Materials Science, Duke University, Durham, NC 27708, USA

* pcsercel@gmail.com

**ABSTRACT**

Recent activity in the area of chiroptical phenomena has been focused on the connection between structural asymmetry, electron spin configuration and light matter interactions in chiral semiconductors. In these systems, spin-splitting phenomena emerge due to inversion symmetry breaking and the presence of extended electronic states, yet the connection to chiroptical phenomena is lacking. Here, we develop an analytical effective mass model of chiral excitons, parameterized by density functional theory. The model accounts for parity mixing of the band edge Bloch functions resulting from polar distortions, resulting in magnetic dipole allowed transitions. Through the study of a prototypical chiral 2D hybrid perovskite semiconductor, we show that circular dichroism of the chiral exciton and its interband continuum emerges from spin-splitting via cross coupling of Rashba-like and chiral/helical spin-texture components. As a counterpoint, we apply our model to describe chiroptical properties of excitons in perovskite nanocrystals that occur without chiral lattice distortions.

A key motivator for developing hybrid semiconductors, i.e., organic/inorganic crystalline systems, is the potential to combine distinct functionalities available in the inorganic layer, e.g., large spin-orbit coupling (SOC), semiconducting or metallic properties, efficient light emission and absorption with tunable energy ranges, ferro- or antiferromagnetism, ferroelectricity, etc., with those available from organic molecules, such as their soft nature, structural diversity, and enantiopure chirality[1]. Chiroptical effects such as circular dichroism (CD)[2-5] and circular polarized light (CPL) emission[6,7] are key examples of physical properties realizable via such combined functionalities; chiral induced spin selectivity (CISS) in transport is another[8]. In many of these hybrid systems the chiroptoelectric properties are found or are assumed to be associated with the spin splitting and resulting spin-textures that occur within the frontier bands, which are derived from the inorganic sublattice rather than the chiral organic molecules. This observation raises the question: What is the mechanism of chiroptic effects, especially at the exciton line, and is there a causal connection to the spin splitting and spin textures?

Here we investigate this question, focusing on the exciton transitions in the chiral 2D layered perovskites *R/S*-NPB (where NPB denotes 1-(1-naphthyl)ethylammonium lead bromide). We focus on these compounds since single crystal x-ray diffraction studies have shown that they possess a relatively simple crystal structure, adopting the non-centrosymmetric space group $P2_1$ with a single inorganic layer per unit cell in contrast to systems of $P2_12_12_1$ symmetry [4]. Moreover, CD has been measured in thin films of *R/S*-NPB; the band structure, exhibiting large spin splitting, has been extensively characterized[4]; and the excitonic CD has been modelled recently using *ab initio* methods[9], providing points of comparison for our mechanistic study.

A great deal has been learned about the mechanisms of structural chirality transfer from the organics to the inorganic constituents in 2D hybrid organic inorganic perovskite semiconductors (HOIPs).[4,5] Chiral organic molecules induce polar distortions in the inorganic framework via asymmetric hydrogen-bonding interactions between the ammonium group of the chiral organic and the halide anions.[3] We can estimate the degree of polar distortion within the chiral structure of *R/S*-NPB by computing the net local formal dipole per unit cell, found for each Pb site from the relative coordinates of the surrounding nearest six Br atoms, taking their charge to be the formal charge of minus *e* (**Extended Table- 1**, **Supplementary Tables S-1, S-2**). In *S*-NPB we find a net local formal dipole directed along the in-plane ***b*** direction, parallel to the $2_1$ screw axis of space group $P2_1$; the *R* enantiomer has a dipole of the same magnitude but oriented oppositely (**Fig. 1 a,b**), while the racemic NPB has no such dipole. The corresponding polarization averaged over the volume of the two PbBr$_6$ octahedra per unit cell is significant at 4.9 $\mu$C/cm$^2$ (**Supplementary Sec. 1**). The dipole direction reversal between *R/S*-NPB is connected with a reversal of the spin textures (**Fig. 1**) and is empirically correlated with a reversal of the CD polarity.[4]

Given this correlation it has been previously suggested that there is direct link between Rashba-like spin splitting and excitonic CD in 2D chiral metal halide perovskites. The simplest such linkage that has been proposed is that excitonic CD is connected to spin splitting via an effective magnetic CD mechanism acting within the electric dipole approximation.[10,11,12] However, CD spectra of 2D HOIPs are typically measured at normal incidence to oriented films and at room temperature. Analysis of the band structure shows that for the out-of-plane direction the bands are very flat (**Supplementary Fig. S-3)** and are consequently prone to localization. Moreover, in the out-of-plane direction spin splitting reverses sign between zone center and edge. Even assuming a band picture for out-of-plane dispersion is appropriate, spin splitting at the nearly degenerate zone center $\Gamma$ and edge Z tends to cancel, resulting in near cancellation of CD response. Similarly, because the spin textures reverse for opposite directions of in-plane wave vector, we can rule out in-plane spin splitting as a direct cause of chiroptical effects measured at normal incidence.[12] Going beyond the electric dipole approximation, a recent computational study made across a series of 2D HOIPs of $P2_12_12_1$ symmetry reported non-zero CD for interband transitions without including SOC, and no correlation between spin splitting energy and CD when SOC is included.[13] A refinement reported in Ref. [9] used *ab initio* methods (GW plus the Bethe Salpeter equation) to examine the chiroptical properties of excitons in chiral S-NPB with SOC. The authors suggested that electron-hole exchange interactions break the degeneracy of the oppositely circularly polarized transitions between Rashba-Dresselhaus-split bands at positive and negative values of in-plane wave vector, leading to CD with a derivative-line-shape. However, neither study considered the full spin textures, which we show are essential to understanding the interplay between spin, excitonic fine structure and CD.

Here, we re-examine the emergence of optical activity within the more flexible framework of a multiband K.P effective mass theory as parameterized by density functional theory (DFT) calculations with SOC. We account for parity mixing of the band edge Bloch functions resulting from polar distortions and include electron hole exchange as well as the full effects of spin-splitting/spin textures via an effective exchange interaction.[14-17] We develop analytical expressions for the exciton fine structure, the electric and magnetic dipole transition matrix elements, and show a direct connection between *particular* spin textures that are unique to chiral semiconductors and the emergence of spin-related CD both for excitonic and continuum interband transitions. As a counterpoint and to demonstrate the generality or our approach, we discuss the mechanisms by which chiroptical properties can occur in perovskite NCs with[18] or in certain instances *without,* polar lattice distortions, and demonstrate that in such three-dimensionally confined systems, chiroptical effects *do not* require spin splitting.

**K.P/Effective mass model.**

**Figure 1** shows the in-plane frontier band dispersion and spin textures in chiral NPB calculated using DFT. The energy splitting (**Fig. 1 c,e**) and corresponding spin textures (**Fig. 1 d,f**) can be understood using the theory of invariants.[19] In general, linear-in-k SOC corrections due to bulk inversion asymmetry (BIA) for dispersion of a given band (valence or conduction) can be written,[5,20] $H_{BIA} = \sum_{i,j} \alpha_{ij} \tau_i k_j$, where $\alpha_{ij}$ are SOC coefficients producing spin polarization with component $i$ for wave vector component $k_j$; the terms, $\tau_i$ are Pauli operators representing total angular momentum, with coordinates $x, y$ in the plane of the inorganic layer and $z$ out-of-plane (**Fig.1c,e**; inset). In a system with space group $P2_1$ ($C_2$ point symmetry) with the $2_1$ screw axis oriented in-plane along the y direction, the symmetry allowed terms for in-plane dispersion are (**Supplementary Sec. 2**),

$$H_{BIA} = \alpha_{zx}\, \tau_z\, k_x + \alpha_{xx}\, \tau_x\, k_x + \alpha_{yy}\, \tau_y\, k_y. \tag{1}$$

If the point symmetry were the non-chiral point group $C_{2v}$, the only allowed non-zero coefficient would be $\alpha_{zx}$, which occurs in non-chiral systems and produces a non-helical "Rashba-like" spin texture for which the spin expectation value is strictly normal to the wavevector $\boldsymbol{k}$.[5] Coefficients with repeated indices, $\alpha_{xx}$ and $\alpha_{yy}$, are allowed in chiral *S/R* NPB by the absence of mirror symmetry. They produce spin textures with *helicity*, i.e., the spin expectation value at wavevector $\boldsymbol{k}$ has components parallel/anti-parallel to the wavevector, leading to spin polarization with components directed in-plane (**Fig. 1d,f**); we characterize these terms as "chiral". Analysis of the band dispersion and spin textures allows the SOC coefficients, effective masses, and parameters determining the structure of the Bloch functions to be fully determined (**Methods** and **Supplementary Sec. 2**), allowing parameterization of an effective mass /K.P model for chiral excitons.

We model the chiral exciton treating the layered 2D HOIP as a set of uncoupled quantum wells with impenetrable barriers, with a dielectric mismatch between the inorganic quantum well and the organic barrier regions. Correspondingly, 2D hydrogenic envelope functions for the exciton are determined variationally, accounting for image charge effects[21] (**Methods**). The exciton binding energy and radius (parameters in Extended Table- 2), are close to results calculated using *ab-initio* methods.[9]

For spin-related mechanisms of CD, it is essential to model the spin-dependent exciton fine structure (EFS). Starting conceptually with a centrosymmetric cubic perovskite, the EFS consists of a lower energy dark singlet state $D$ split from an upper bright triplet due to short range (SR) exchange (**Figure 2**).  The degeneracy of the bright triplet levels is broken in a centrosymmetric 2D HOIP by the effects of anisotropic confinement and in-plane anisotropy. This is captured via phenomenological "tetragonal" crystal field (CF), $\delta$,[22] and "orthorhombic" CF[15], $\zeta$, resulting in a lower $Z$ level and two higher energy levels, $Y, X,$ split by the orthorhombic CF. The electric dipole (ED) transition moments are oriented out-of-plane ($Z$) or polarized along the in-plane ***b*** direction ($Y$) and ***a*** direction ($X$). Long-range (LR) exchange corrections[23,24] shift the $Z$ state above the $Y, X$ pair as verified experimentally for phenethylammonium lead iodide[25] and via *ab initio* calculations in model layered perovskites[26]. Relative to a centrosymmetric system such as racemic-NPB, the EFS in chiral R/S-NPB is further modified by the structural chirality transfer in two ways. First, considering the polar distortion as a perturbation on a reference centrosymmetric structure, the even and odd parity valence and conduction band Bloch functions mix (**Methods**).  This causes small modifications in the SR and LR exchange energies, but more importantly, leads to allowed magnetic dipole (MD) and electric quadruple (QE) transitions, whose magnitude is proportional to the polar distortion and which would otherwise be forbidden.   Second, energy shifts, and critically, level mixing, occur due to the electron and hole spin splitting, which lead in second order perturbation theory to an effective exchange interaction. Generalization of the "Rashba exciton" theory[14-17] to the chiral excitonic system of point symmetry $C_2$, with polar distortion along the in-plane $y$ direction,  shows first, that levels $D, Z$ shift up by an amount $\Delta_{zz}$, and levels $X, Y$ shift down by $-\Delta_{zz}$, where $\Delta_{zz} \cong -2A_{rel} \frac{\mu}{\hbar^2}(\alpha_{zx}^e \alpha_{zx}^h)$ is due primarily to the dominant electron and hole Rashba SOC terms $\alpha_{zx}^{e(h)}$; $A_{rel}$ is a numerical factor determined by the material parameters (**Methods**, and Eq. S3.60) and $\mu = (1/m_e + 1/m_h)^{-1}$ is the in-plane exciton reduced effective mass. Second, the chiral terms $\alpha_{xx}^{e(h)}$ couple with the non-chiral Rashba-like terms $\alpha_{xz}^{h(e)}$, leading to pairwise mixing of exciton levels $D$ & $Y$ and $X$ & $Z$, due to the generation of coupling terms $\Delta_{DY}, \Delta_{XZ}$ given by,

$$\Delta_{DY} = -2A_{rel}\frac{\mu}{\hbar^2}(\alpha_{zx}^e\alpha_{xx}^h - \alpha_{xx}^e\alpha_{zx}^h), \quad \Delta_{XZ} = -2A_{rel}\frac{\mu}{\hbar^2}(\alpha_{zx}^e\alpha_{xx}^h + \alpha_{xx}^e\alpha_{zx}^h). \quad (2)$$

(See **Methods** and **Supplementary Sec. 3**). This level mixing is essential for generation of chiroptical effects since the rotary strength vanishes if the electric and magnetic transition dipoles of each fine structure level are orthogonal, which is the case in a system of $C_{2v}$ symmetry with vanishing chiral terms $\alpha_{xx}^{e(h)}$ (**Fig. 2b**; **Supplementary Table S-9** for symmetry analysis). **Figure 2c** tabulates the EFS energies for chiral S-NPB with the corresponding electric and magnetic transition dipoles. Analytical expressions for the transition moments are summarized in **Extended Table- 3**; symmetry analysis in **Supplementary Table S-10**. **Figure 2c** also shows the relative rotary strength, $\mathcal{R}_{CD,\perp}$, (Eq. S3.71) and dissymmetry, $g_{CD,\perp}$, (Eq. S4.15), for illumination by light incident normal to the inorganic layers. The pairwise mixing, Eq. 2, ensures that the excitonic CD shows a derivative-like Cotton-effect (**Extended Table- 4** for analytical expressions). We note that our numerical calculations include only the effect of coupling between the electric and magnetic transition dipoles. Unlike the magnetic transition dipoles, inter-band electric quadrupole transition matrix elements involve significant contributions from remote bands that cannot be reliably estimated within a near-band K.P framework. Nevertheless, symmetry considerations allow assessment of the pairwise mixing on quadrupole contributions, summarized in **Extended Table- 4**.

**Calculation of CD spectra**

With the EFS determined, we compute the relative dielectric tensor response of the chiral exciton for light incident normal to the inorganic layers, including magnetic dipole corrections (Eq. M14, and **Supplementary Sec 4**). Starting from the wave equation we use biorthogonality relations to perform a normal mode decomposition (**Methods**), computing the decadic absorbance $\mathbb{A}$ of incident light of left- and right- circular polarization and the resulting CD. The CD (**Fig. 3b**) exhibits a Cotton effect, with polarity reversal between the two enantiomers. The polarity reversal tracks a reversal of the orientation of the magnetic transition dipole of each fine structure level for the two enantiomers relative to the corresponding electric transition dipoles (**Extended Table- 5**) which is due to the reversal of the polar distortion between chiral *S*-NPB and *R*-NPB (**Fig. 3 c,d**). We compare measured frontside/backside average CD and absorbance measured for thin film S-NPB against the model, matched to the experimental optical thickness and linewidth (**Fig. 4, Methods**). The averaging is performed to mitigate "apparent CD" effects[27,28] which can occur due to thin film morphology[29]. Both the calculated and measured CD show a Cotton effect of the same polarity; the calculated CD range (maximum – minimum) is 5.2 millidegree is slightly smaller than the measured range of 6.4 millidegree possibly due to the neglect of electric quadrupole contributions, which also exhibit a Cotton effect.

We next explore the connection between CD and spin textures (**Fig. 5**). **Fig. 5a** shows conduction band dispersion and spin textures calculated with all terms from Eq. 1, while **Fig. 5 b**, **c** respectively show the dispersion calculated retaining only the non-chiral Rashba-like $\alpha_{zx}^{e(h)}$ or only the chiral $\alpha_{xx}^{e(h)}, \alpha_{yy}^{e(h)}$ SOC terms. The corresponding absorbance and CD are displayed in **Fig. 5e-f**. Without both the Rashba *and* the chiral SOC terms which cause mixing of the fine structure levels (Eq. 2), excitonic CD vanishes. The same condition holds as well for the continuum interband (IB) transitions (**Methods**). Notably, IB rotary strengths (**Extended Table- 6**) are independent of the spin-splitting energies in either band.

**Generalization to perovskite nanocrystals**

A recent study demonstrated that surface-adsorbed chiral ligands can induce inversion-symmetry-breaking distortions in chiral-ligated $CsPbBr_3$ nanocrystals, and suggested that this is the likely cause of observed CD.[6] Given this result and the recent prediction by Swift *et al*. of ferroelectric phases in metal halide perovskite nanocrystals that exhibit Rashba-like spin splitting[18], we generalized the exciton model to describe perovskite NC systems with polar distortions (**Methods**). To avoid the complexity presented by the spatial non-uniformity of polar distortions in chiral-ligated NCs[6], we applied the model to address the possibility of chiroptical effects in colloidal ferro-electric $CsPbBr_3$ nanocrystals (**Methods**). Importantly, while the bulk ferro-electric $CsPbBr_3$ crystal structure predicted by Swift et al. is non-chiral with point group symmetry $C_s$, in NC form, the pseudo-cubic bounding facets[30], which are misaligned to the primitive lattice vectors, can break the mirror symmetry, creating a chiral structure. This is illustrated in **Figure 6**. Panels **a-d** show that while ferro-electric NCs with perfect cube shape (equal basal edge lengths $L_x=L_y$) have non-chiral symmetry $C_s$ and therefore exhibit zero CD, NCs with unequal basal edge lengths have chiral symmetry $C_1$ and exhibit weak CD in orientationally random colloids (panel **b**). Significantly, the polarity of the CD response reverses if we reverse the basal edge length ratio (**Fig. 6 b**, **Extended Table- 8**): It is the interplay between LR exchange and the NC shape[30], not Rashba exciton effects, which leads to the fine structure level mixing that enables CD. Indeed, if we artificially toggle the Rashba effect off, setting the exciton Rashba energy $\mathcal{E}_R$ to zero, (**Methods**, Eq. M19), the absorbance and CD are practically unaffected (open symbols in panels **a**, **b**). Conversely, NCs artificially modelled with bounding facets normal to the primitive lattice vectors, which we label in panels **a**, **b**, **e** "quasi-orthorhombic facets", exhibit zero CD since in that case the bounding facets do not lower the point symmetry of the NC from $C_s$. Correspondingly, the rotary strength is zero for each fine structure level (**Extended Table- 9**).

The foregoing analysis assumes orientationally random colloids. Were the NCs crystallographically oriented, the situation would be different. We illustrate this by modelling dense oriented NC arrays, which may become feasible given recent advances in super-assembly techniques.[31,32] In such assemblies, large "apparent CD" effects[27] can arise even in non-polar $CsPbBr_3$ NCs, which is antisymmetric, reversing sign upon reversal of the direction of incident light, and does not exhibit a Cotton effect, as shown in Fig. 6 panels **f**, **g**. The effect is a second consequence of the interplay between LR exchange and NC shape discussed above: In NCs bounded by pseudocubic facets, and with unequal basal edge lengths, LR exchange coupling mixes bright exciton states resulting in non-degenerate, non-orthogonal electric transition dipoles, a condition shown previously[27] to cause excitonic "apparent" CD (**Supplementary Sec 6.4.1**, **Extended Table- 8**, **Extended Table- 10** ). These calculations may provide insight into the nature of large chiroptic effects observed recently in single $CsPbBr_3$ nanocrystals, which exhibit a degree of circular polarization of up to 38% in photoluminescence emission from individual nanocrystals[33], far larger than expected from our calculated dissymmetry factors for intrinsic CD.

**Conclusions**

We have presented a K.P/effective mass theory for chiroptical activity in chiral 2D HOIPs and contrast with the appearance of CD in perovskite NC systems. The model accounts for parity mixing of the band edge Bloch functions resulting from polar distortions, including its effects on electron hole exchange and the electric and magnetic transition dipole matrix elements. The effects of spin-splitting on excitons is included via an effective exchange interaction determined by the particular spin textures of each system. Analytical expressions demonstrate a direct connection between particular spin textures that are unique to chiral semiconductors and the emergence of CD both for excitonic and continuum interband transitions. For the exciton, cross coupling of chiral and non-chiral SOC terms mixes EFS levels in a pairwise fashion, leading to the emergence of excitonic CD with derivative-like line-shape in chiral 2D perovskites. As a counterpoint, we show that the mechanism by which intrinsic chiroptical effects occur in ferroelectric nanocrystals is distinct, proceeding by LR exchange mixing of EFS levels, i.e., spin splitting effects are inessential. We also demonstrate non-reciprocal "apparent CD" effects in oriented assemblies that can arise even in non-polar NCs, which may provide insight into the nature of large chiroptic effects observed recently[33] in single $CsPbBr_3$ nanocrystals.

## Online content

Methods, Extended Tables, Supplementary Information, peer-review information, and statements of data and code availability are presented in the online version of this paper.

# Figures

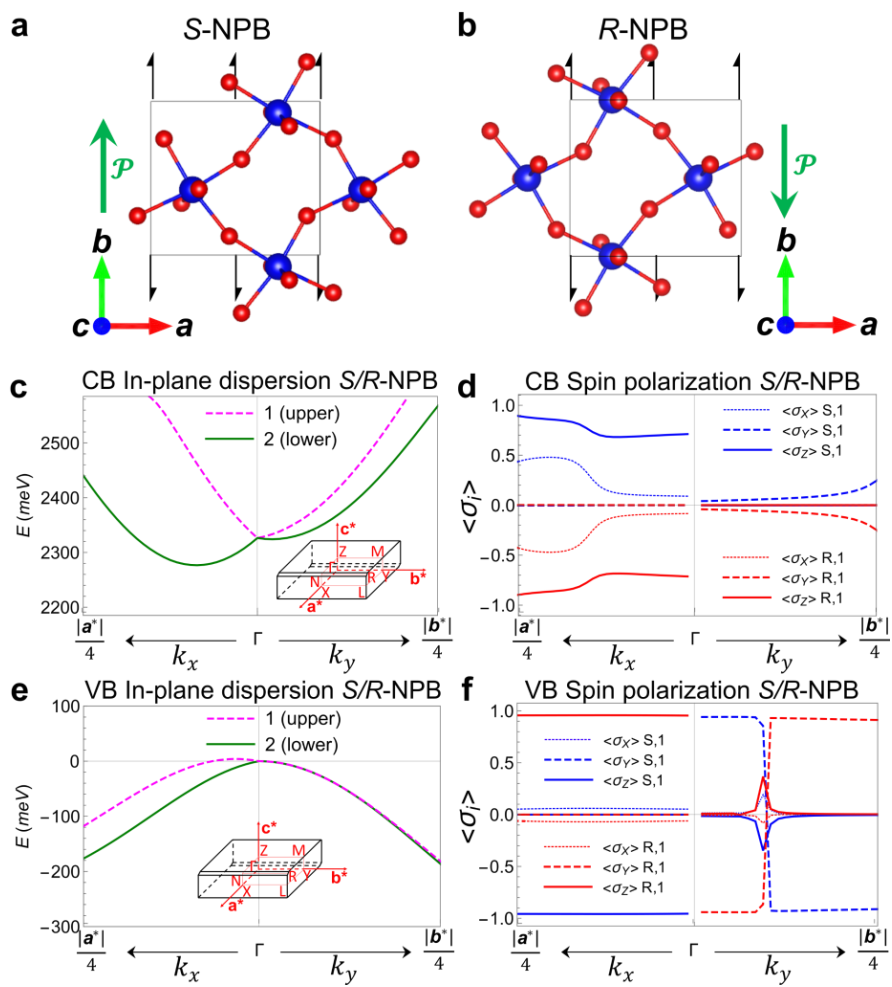

**Figure 1 | In-plane polar distortion, frontier band dispersion and spin textures in chiral NPB.** In-plane view of *S*-NPB, **a**, and *R*-NPB, **b**, crystal packing diagrams based on 298K single crystal x-ray diffraction data.[4] Pb and Br are denoted as blue and red spheres, respectively. The location of the two-fold screw axes are indicated with black $2_1$ arrows parallel to the direction of unit vector ***b*** in all cases. The direction of the average local polarization due to the net local formal dipole at the Pb sites is shown with green arrows labelled $\mathcal{P}$ and is oppositely directed in the S- and R- structures. The local dipole and polarization are zero in the racemic structure (not shown). See **Extended Table- 1** and **Supplementary Tables S-1**, **S-2**. Panel **c**, conduction band (CB) dispersion near Γ in the plane of the inorganic layers. The corresponding spin polarization of the upper sub-band ("1") for *S*/*R*- NPB are shown in panel **d**; panels **e** and **f** show the in-plane valence band (VB) dispersion near Γ and the corresponding upper sub-band spin polarization. Coordinates are chosen such that *x, y* directions align to the primitive vectors ***a, b***, respectively, while the *z*-axis is orthogonal to the plane of the inorganic layers. The higher energy sub-band in all plots is denoted "1" while the lower energy sub-band is denoted "2". Note the spin texture reversal between the S and R enantiomers in panels **d** and **f**, which correlates to the polarization direction reversal in panels **a** and **b**. Calculations were performed using DFT-PBE+SOC on geometrically relaxed structures (**Methods**).

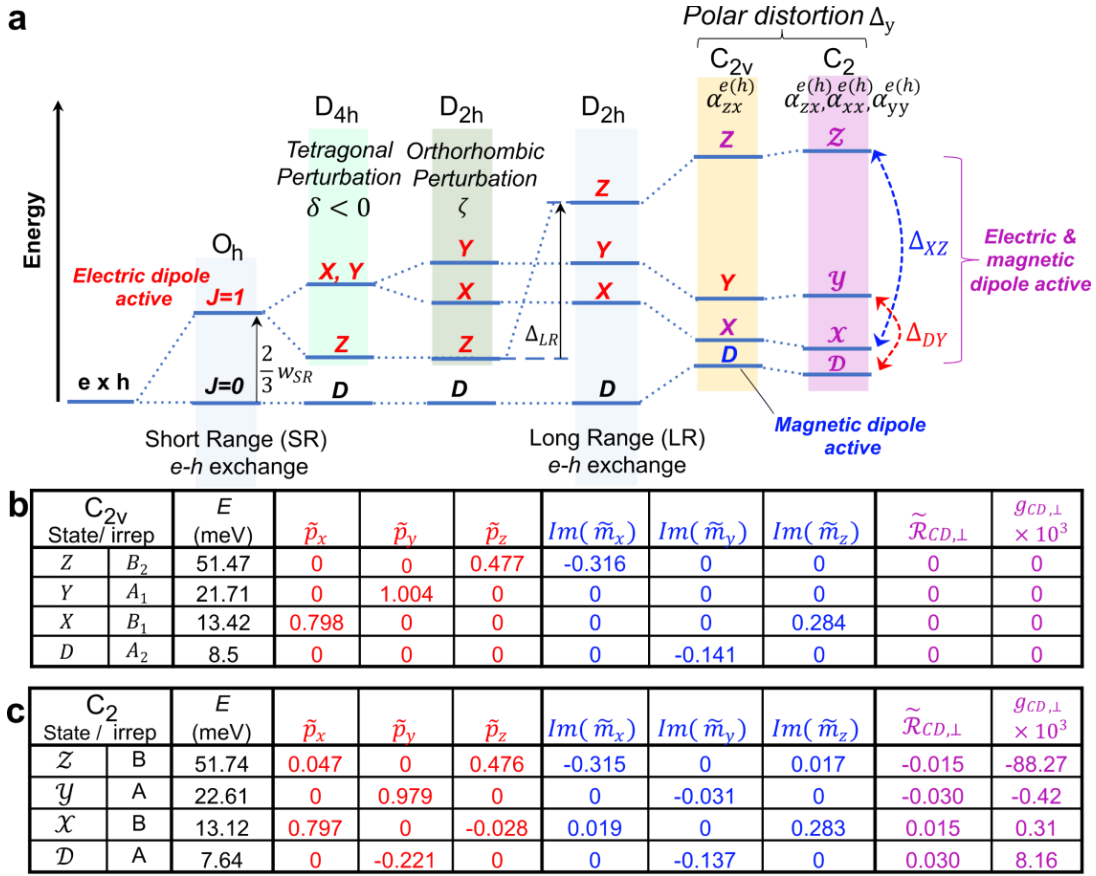

**Figure 2 | Exciton fine structure in *S*-NPB. a** Schematic of the descent in symmetry from point group $O_h$ to $C_2$ of the exciton fine structure created by: The short-range (SR) electron/hole (e/h) exchange interaction; perturbations associated with anisotropic confinement of the inorganic layers, modelled via tetragonal crystal field, $\delta$, and in-plane anisotropy, modelled via "orthorhombic" crystal field, $\zeta$; long-range (LR) e/h exchange interaction; and effective exchange interaction due to Rashba-like splitting associated with polar distortion along in-plane y-axis. SR exchange splits the levels into a lowest energy dark singlet level, $D$, and an upper bright triplet, $X, Y, Z$, with electric dipole transition moments aligned respectively along the *x, y,* and *z* directions. Crystal field $\delta$ shifts $Z$ from $X/Y$, while crystal field $\zeta$ splits the $X$ and $Y$ levels. The LR interaction shifts the $Z$ state up in energy. With polar distortion, magnetic dipole transition moments become allowed. An effective e/h exchange interaction including only non-chiral Rashba-like spin splitting terms, $\alpha_{zx}^{e(h)}$, shifts levels $D, Z$ up; $X, Y$ down; the fine structure for this case is shown in panel **b**. Addition of chiral spin splitting terms, $\alpha_{xx}^{e(h)}$, couples $X, Z$, and $D, Y$, resulting in mixed states labelled $\mathcal{D}, \mathcal{X}, \mathcal{Y},$ and $\mathcal{Z}$, whose fine structure is given in panel **c**. Panels **b, c** show the irreducible representation (irrep) of each level[34] with energy correction (*E*), dimensionless electric ($\tilde{p}$) and magnetic ($\tilde{m}$) transition dipoles (related to dimensioned transition dipoles through $\boldsymbol{P} = i|P_K|\mathcal{K}\,\tilde{\boldsymbol{p}}$, and $\boldsymbol{M} = \hbar\mathcal{K}\tilde{\boldsymbol{m}}$ in terms of common factors $P_K$, the Kane momentum matrix element[35], and $\mathcal{K}$, the overlap factor). Relative rotary strength, $\tilde{\mathcal{R}}_{CD,\perp}$, (Eq. S3.71) and dissymmetry, $g_{CD,\perp}$, (Eq. S4.15) are given for illumination by light incident normal to the inorganic layers. Calculated values use parameters in **Extended Table- 2**. See **Supplementary Section 3** for further details.

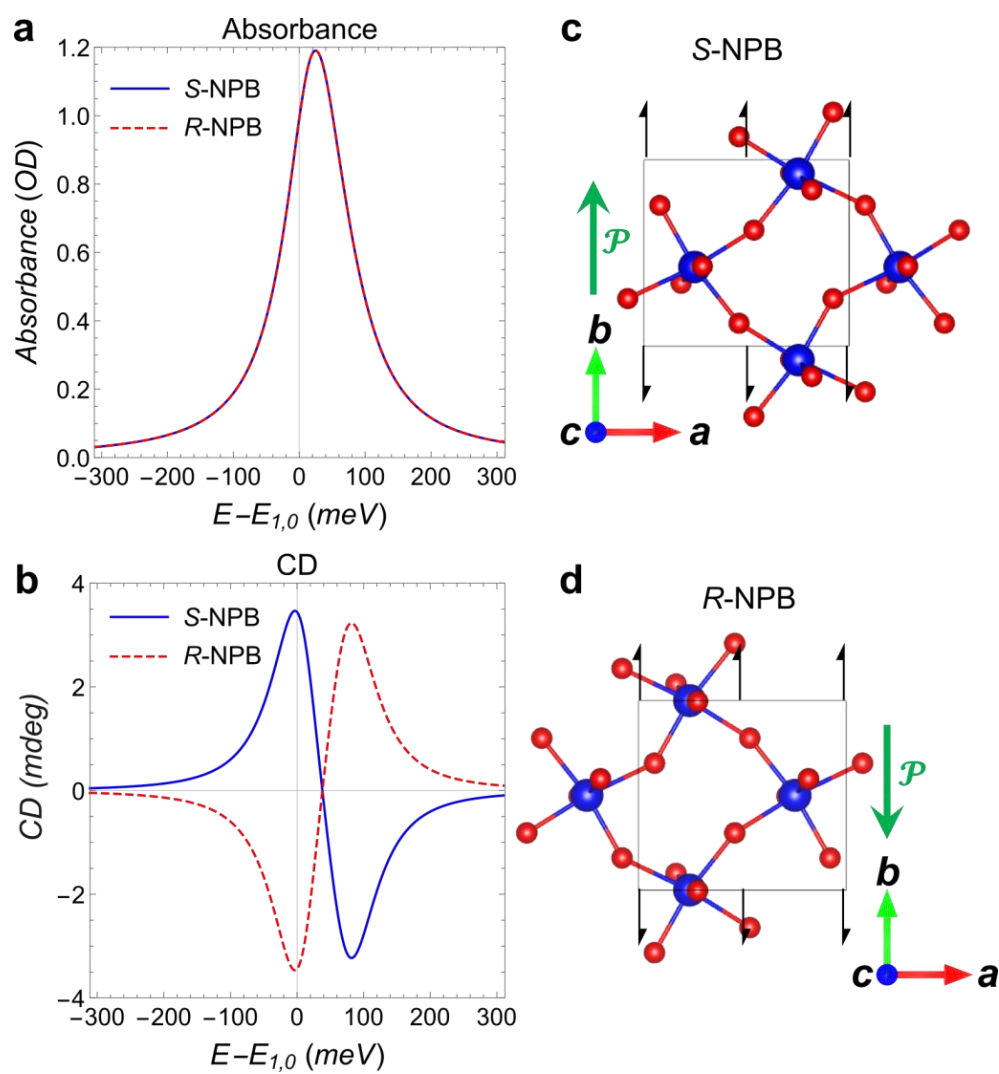

**Figure 3 | CD in chiral NPB. a**, Absorbance in optical density (OD) units and **b**, CD in ellipticity units (mdeg= millidegree), calculated for *S*-NPB, (blue lines), and *R*-NPB (dashed red lines) for model film thickness 120nm, using a Lorentzian line shape function set to realize full width at half maximum absorbance of 115 meV. Spectra are plotted versus energy relative to the energy, $E_{1,0}$, of the lowest 2D exciton with principal quantum number n=1, azimuthal quantum number m=0, in the absence of fine structure corrections. The CD of *S*-NPB and *R*-NPB are opposite in polarity. This polarity reversal is due to a reversal of the direction of the magnetic transition dipole of each fine structure level in relation to the corresponding electric transition dipoles (**Extended Table- 5**). This reversal in turn is due to the oppositely directed polar distortions of chiral *S*-NPB and *R*-NPB as depicted in panels **c** and **d** respectively. Calculations use parameters in **Extended Table- 2**. Spin splitting coefficients in *R*-NPB are equal in magnitude but opposite in sign from those of *S*-NPB as indicated by the spin texture plots shown in **Figure 1**. See **Supplementary Section 1- 4** for further details.

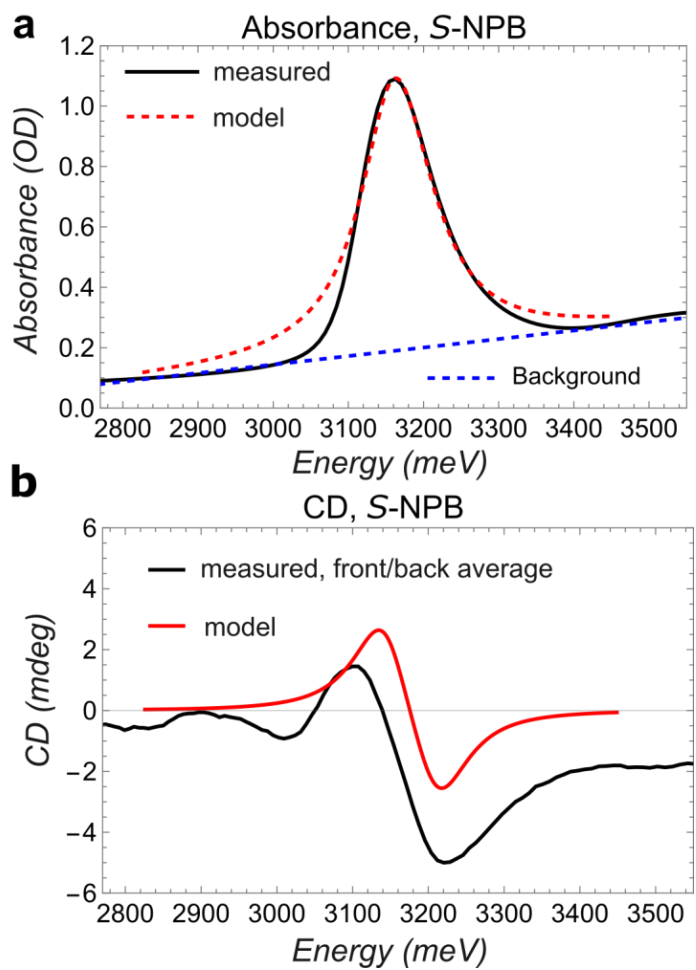

**Figure 4 | Comparison of CD theory and measurement, *S*-NPB films. a**, Absorbance in optical density (OD) units and **b**, CD in ellipticity units (mdeg= millidegree) plotted versus transition energy. Black lines in the plots show the front-side/back-side average of the measured absorbance and CD of a *S*-NPB film to mitigate "apparent CD" effects due to film morphology (see **Supplementary Section 7**) and are compared with theory displayed with red lines. Calculations are made using a Lorentzian line shape function with broadening set to match the measured full width at half maximum (115 meV) of the experimental absorbance spectrum. The model film thickness was adjusted to 90 nm to match the peak absorbance in the measured film after subtracting the background signal (shown as the dashed blue line in panel **a**). Calculations use parameters listed in **Extended Table- 2**. See **Supplementary Sections 1-4** for further details. The experimental absorbance and CD separately measured from the front and back sides of the sample are shown in **Supplementary Figure S-9**.

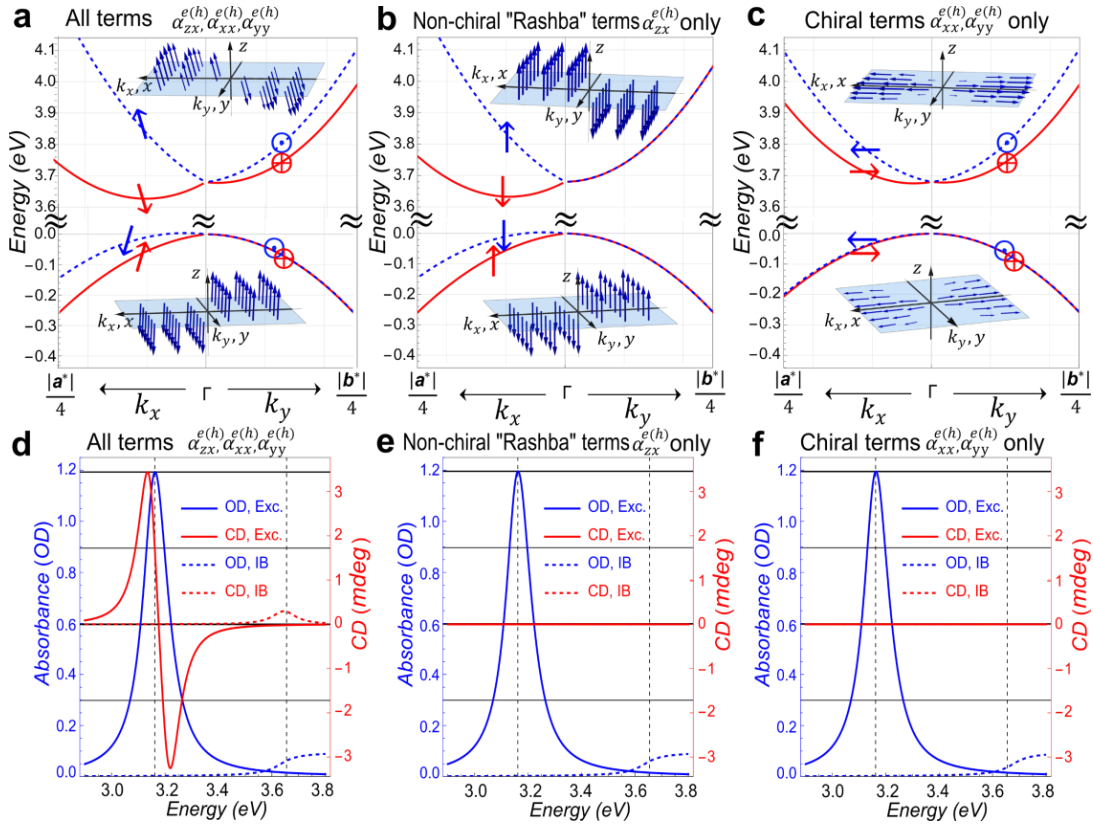

**Figure 5 | Impact of spin texture on CD.** Panel **a** shows in-plane conduction and valence band dispersion and spin textures for *S*-NPB calculated within K.P theory. Spin textures for non-degenerate sub-bands are indicated schematically with colored arrows, with *y* out-of-page; insets show full spin texture of uppermost sub-bands, with in-plane components *x, y* aligned to $k_x, k_y$ and magnified by 2 ×; the *z*-axis is orthogonal to the plane of the inorganic layers. The corresponding absorbance (blue lines, optical density "OD" units) and CD (red lines, ellipticity units,"mdeg"=millidegree) are shown in panel **d**. Spectra are plotted versus transition energy and display the response of the lowest exciton ("Exc.", solid lines) as well as the interband (IB) continuum component (dashed lines) which includes no e/h interactions. For comparison, panels **b** and **e** respectively show the in-plane band dispersion/spin textures and absorbance /CD computed by including only non-chiral Rashba-like spin splitting terms, $\alpha_{zx}^{e(h)}$, with chiral spin splitting terms set to zero: $\alpha_{xx}^{e(h)} = \alpha_{yy}^{e(h)} = 0$. Spin polarization for this case is orthogonal to the plane of the inorganic layers. Panels **c** and **f** correspondingly show the band dispersion/spin textures and absorbance / CD with the chiral spin splitting terms, $\alpha_{xx}^{e(h)}$, $\alpha_{yy}^{e(h)}$ non-zero, and non-chiral "Rashba" terms set to zero: $\alpha_{zx}^{e(h)} = 0$. For this case, the spin polarization is always directed in the plane of the inorganic layers. The calculations demonstrate that non-zero CD can only occur when both the non-chiral ($\alpha_{zx}^{e(h)}$) and chiral ($\alpha_{xx}^{e(h)}$) spin splitting terms are present, both for the exciton and interband continuum transitions. Calculations use parameters listed in **Extended Table- 2**. Exciton fine structure energy corrections, transition dipole matrix elements, and rotary strength for each case shown are given in **Supplementary Table S-14**. For analytical expressions for the rotary strength of the excitonic and IB transitions see **Extended Table- 4**, **Extended Table- 6.**

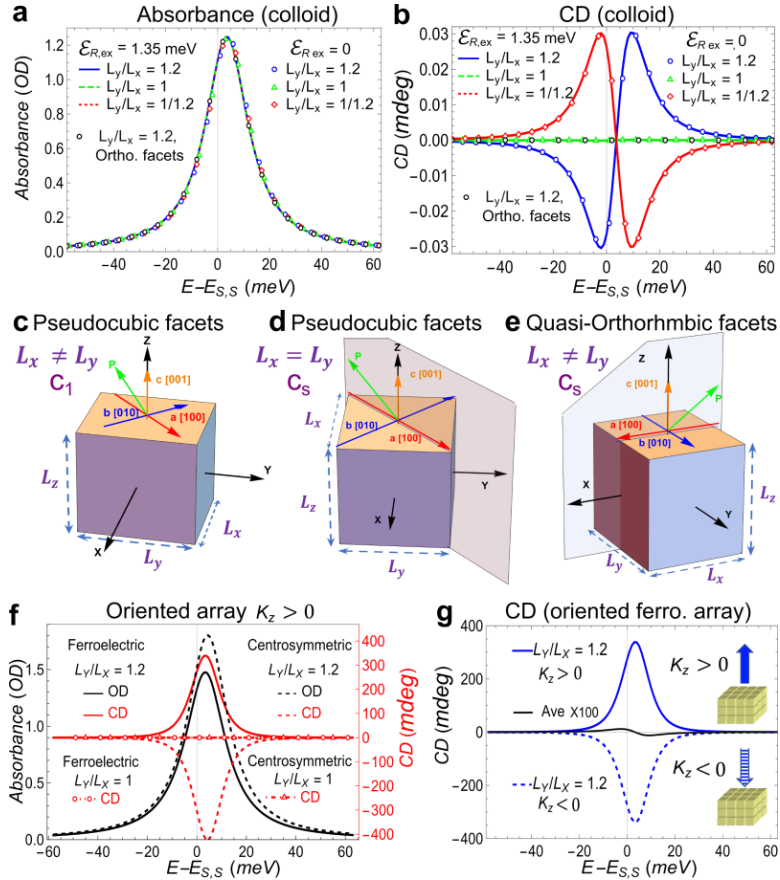

**Figure 6 | CD in CsPbBr$_3$ nanocrystals.** Panels **a**,**b**: Absorbance in optical density (OD) units and CD in ellipticity units (mdeg=millidegree) for colloidal ferroelectric NCs of size $L_e = \sqrt[3]{L_x L_y L_z} = 10$ nm, colloid volume fraction 0.001 and sample path length 1mm. Transition linewidth is 20 meV. Spectra are plotted versus energy relative to the energy, $E_{S,S}$, of the lowest exciton in the absence of fine structure corrections. Blue, green and red lines correspond to ferroelectric NCs with pseudocubic facets, with basal edge length ratios $L_y/L_x = 1.2$, 1 and 0.83, respectively. Panel **c**, **d**: Model of ferroelectric NC with unequal versus equal basal edge lengths, point groups C$_1$, C$_s$ respectively. Green arrows depict orientation of the polar distortion (see **Supplementary Table S-14**); mirror planes are shaded. In panel **b**, NCs with $L_y/L_x = 1$ are non-chiral with CD=0, while for unequal basal edge length ratio the CD polarity depends on $L_y/L_x$. Panels **a**,**b** also show calculations for ferroelectric NCs with exciton Rashba energy $\mathcal{E}_R = 0$, with $L_y/L_x = 1.2$, 1 and 0.83, corresponding to blue circles, green triangles, and red diamonds. Black circles indicate ferroelectric NCs with quasi-orthorhombic bounding facets (panel **e**) with $L_y/L_x = 1.2$, a nonchiral structure (C$_s$; mirror plane is shaded), exhibiting no CD. Panel **f** shows absorbance and CD spectra for dense oriented arrays of ferroelectric and centrosymmetric NCs with pseudocubic bounding facets with the ***c***-axis along the positive z-direction, taken to be the optical axis. For unequal basal edge lengths, CD is dominated by "apparent CD" which is monopolar and antisymmetric with respect to the direction of incident light as shown in panel **g** for ferroelectric NCs. Directionally averaged CD shows a Cotton-effect for ferroelectric NCs but vanishes for centrosymmetric NCs (**Supplementary Fig. S-6**). Parameters used are listed in **Extended Table-7** unless otherwise specified.

## Methods

### Materials

(*S*)-(−)-1-(1-naphthyl)ethylamine (≥99%, Sigma Aldrich), lead bromide (PbBr2, ≥98%, Sigma Aldrichchemicals), hydrobromic acid (HBr) (48 wt% in H2O, >99.99%, Sigma Aldrich), Diethyl ether (Anhydrous (≥99, Fisher Chemical, stabilized with BHT), 2-methoxyethanol (anhydrous, 99.8%, Sigma Aldrich), and were procured and used without further purification.

### Synthesis

Single crystals of (S-NEA)$_2$PbBr$_4$ were grown following a previous method.[36] In brief, lead (II) bromide (90 mg, 0.24 mmol) was dissolved in 1.0 mL of concentrated hydrobromic acid by heating at 90°C for <5 minutes. The solution was cooled to room temperature and 2.4 mL DI water was slowly added. (S)-(-)-1-(1-napthyl)ethylamine (78 µL, 0.48 mmol) was then added and the mixture was dissolved in a sealed vial at 90°C overnight while stirring. The solution was cooled to room temperature at a rate of 1°C/h in an aluminum heating block. The (S-NEA)$_2$PbBr$_4$ crystals were collected via vacuum filtrations and the crystals were washed with 20 mL of diethyl ether three times. The crystals were then dried overnight in a vacuum oven at 55°C, 150 Torr.

### Film characterization

To prepare films for CD and absorption experiment, glass microscope slides were cut into 1x1 in$^2$ pieces and cleaned by sonicating in isopropyl alcohol for 15 minutes followed by UV-ozone treatment for 15 minutes. In a nitrogen atmosphere, the collected single crystals of (S-NEA)$_2$PbBr$_4$ were dissolved in 2-methoxyethanol (100 mg/mL) followed by filtering through a 0.22 µm PTFE filter. The solution (100 µL) was spin coated on the cleaned glass substrates at 4000 RPM for 20 s followed by annealing at 120°C for 10 minutes. CD spectra were collected on an Olis DSM spectropolarimeter at 1 nm interval with 10s integration time. Front and back

CD spectra (**Supplementary Section 7** and **Supplementary Fig. S-9**) were collected in the same position on the film. Absorption spectra were collected on a Cary 7000i spectrometer.

**First-principles calculations**

First-principles density functional theory calculations were carried out using FHI-aims[37] with the numeric atom-centered orbital (NAO) basis sets. For these calculations, experimental structures based on single-crystal XRD measurements were taken from the literature as follows: *S*-NPB, Ref.[5], *R*-NPB [4] and rac-NPB Ref. [4]. To obtain relaxed geometries, structures from the XRD experiments went through a geometry optimization process based on the semilocal PBE functional[38] plus the TS pairwise dispersion scheme for van der Waals (vdW) interactions.[39] The electronic structure of both experimental and relaxed geometries were calculated based on the PBE functional and second variational perturbative spin-orbit coupling.[40] Spin-texture calculations were carried out as described in Ref. [4]. Along each selected reciprocal path, 51 points were evenly sampled to depict the band structure and spin textures.

**K.P and effective-mass-approximation calculations.**

Analysis of spin textures and modelling of exciton level structure was performed using the Mathematica v.14 programming environment, using custom-written software.

**Effective mass model for 2D excitons with polar distortion**:

The wave function of a quasi-2D exciton in a chiral 2D HOIP can be written in the effective mass approximation as a product wave function in the form of a Bloch wave,

$$\psi^{ex}_{e_i,h_j K;nm}(\boldsymbol{r}_e, \boldsymbol{r}_h) = w^e_i(\boldsymbol{r}_e)\, w^h_j(\boldsymbol{r}_h)\, f_{K;nm}(\boldsymbol{r}_e, \boldsymbol{r}_h). \tag{M1}$$

Here, $f_{K;nm}(r_e, r_h)$ represents an envelope wavefunction characterized by wave vector $K$ with relative electron/hole motion described by a 2D hydrogenic function with principal and azimuthal quantum numbers $n, m$. Terms $w_i^e$ and $w_j^h$ denote band-edge periodic Bloch functions for the electron and hole. Exciton binding energies and corresponding exciton envelope functions are computed using a variational procedure which accounts for the dielectric discontinuity between inorganic well and organic barrier layers via an image charge potential as developed in Refs. [21] and detailed in **Supplementary Sec 3.1**. Model parameters are summarized in **Extended Table- 2**. For calculations of the exciton fine structure and optical properties in chiral perovskites, we account for the effect of polar distortions on the periodic Bloch functions using perturbation theory. We start with racemic-NPB as a reference unperturbed non-polar structure with bandgap $E_{g,NP}$ ("NP" denoting non-polar) and band-edge Bloch functions $u_1^v, u_2^v, u_1^c, u_2^c$ of definite parity given by[15,30],

$$u_1^v = s \uparrow, \qquad\qquad u_2^v = s \downarrow,$$
$$u_1^c = -\mathcal{C}_Z\, p_Z \uparrow -(\mathcal{C}_X\, p_X + i\mathcal{C}_Y\, p_Y) \downarrow, \quad u_2^c = -(\mathcal{C}_X\, p_X - i\, \mathcal{C}_Y\, p_y) \uparrow + \mathcal{C}_Z\, p_Z \downarrow. \tag{M2}$$

Here, the term symbol *s* denotes a periodic "orbital" function that transforms as an invariant under the operations of the crystal point symmetry group; $p_X$, $p_Y$, and $p_Z$ similarly denote periodic functions that transform like *x ,y, z*; ↑ (↓) denote the spin functions with projection $S_z = +1/2\ (-1/2)$; while $\mathcal{C}_X, \mathcal{C}_Y, \mathcal{C}_Z$ are real numbers expressible in terms of phase angles, $\theta, \phi$, that reflect the effect of crystal field splitting (**Supplementary Sec.1**):

$$\mathcal{C}_X \approx \frac{\cos\phi \cos\theta - \sin\phi}{\sqrt{2}}\ , \mathcal{C}_Y = \approx \frac{\cos\phi \cos\theta + \sin\phi}{\sqrt{2}}\ , \mathcal{C}_Z \approx \cos\phi \sin\theta. \tag{M3}$$

In these expressions, the phase angle $\theta$ is determined by the spin orbit split-off parameter, Δ, and the "tetragonal" crystal field[22] $\delta$ while the phase angle $\varphi$ reflects in-plane anisotropy and is determined by "orthorhombic" crystal field[15] $\zeta$:

$$\tan 2\theta = \frac{2\sqrt{2}\,\Delta}{\Delta - 3\delta}, \quad \tan 2\varphi = \frac{-4\,\zeta\,\cos\theta}{\Delta + \delta + \sqrt{\Delta^2 - \frac{2}{3}\Delta\,\delta + \delta^2}}\,;\ \theta \leq \frac{\pi}{2}\,. \tag{M4}$$

In the presence of a polar distortion, due to the accompanying polarization, $\mathcal{P}$, the parity eigenfunctions in Eq. M2 mix, resulting in a band gap increase and mixed parity band edge Bloch functions. For chiral $S$-NPB, analysis of the net local formal dipole about the Pb atoms (**Supplementary Sec 1.2.1**) shows that the polarization is directed along the primitive vector ***b*** axis, parallel to the $2_1$ screw axes of the experimental structure, defined as the $+y$ direction; analysis of $R$-NPB shows that the polarization is oppositely directed while the polarization in rac-NPB is zero. The band gap of the polar structure is given approximately by $E_g \approx E_{g,NP}(1 + 2\delta_Y^2)$, where $\delta_Y$ is a parity mixing amplitude, given for small distortion by $\delta_Y \sim \mathcal{C}_y \Delta_y / E_{g,NP}$, where perturbation potential $\Delta_y$ is proportional to the polarization along $y$. The mixed parity band edge Bloch functions $w_1^v, w_2^v, w_1^c, w_2^c$ are then given in the $u_1^v, u_2^v, u_1^c, u_2^c$ basis by,

$$w_1^v = \frac{1}{\sqrt{N_w}}(u_1^v + i\delta_Y u_2^c),\quad w_2^v = \frac{1}{\sqrt{N_w}}(u_2^v - i\delta_Y u_1^c);$$
$$w_1^c = \frac{1}{\sqrt{N_w}}(-i\delta_Y u_2^v + u_1^c),\quad w_2^c = \frac{1}{\sqrt{N_w}}(i\delta_Y u_1^v + u_2^c), \tag{M5}$$

where $N_w$ is a normalization factor. (Expressions for a polar distortion of general orientation are given in **Supplementary Sec. 1.2**). Importantly, since the bandgap increases with polar distortion relative to $E_{g,NP}$, the bandgap shift is used as a constraint to determine the polarization perturbation potential $\Delta_Y$ in the chiral $S$-NPB structure concurrent with analysis of the spin textures, which is required to determine the crystal field parameters in addition to the SOC coupling coefficients (**Supplementary Sec 2**). The parity mixing reflected in Eq. M5 leads to magnetic dipole and electric quadrupole transition matrix elements whose magnitude is proportional to the parity mixing amplitude $\delta_Y$, see supplementary Eq. S3.47 and the expressions in **Extended Table- 3**.

**Analysis of spin textures.**

To determine spin-splitting coefficients in chiral *S*-NPB, we generalize the approach developed in Refs. [5,20] to explicitly account for the parity mixing of the Bloch functions, Eq. M5. In brief: Energy splitting due to Rashba-like SOC along a given direction in reciprocal space, denoted $k_j$, where index $j$ runs over $x, y, z$, is determined by an effective SOC coefficient, $\alpha_{eff,j}$ given by,

$$\Delta E(k_j) = \pm \sqrt{\sum_i \alpha_{ij}^2} \, k_j \equiv \pm \alpha_{eff,j} k_j, \tag{M6}$$

where index $i$ runs over $x, y, z$, and $\alpha_{ij}$ are individual symmetry allowed coefficients producing spin polarization with component direction $i$ for wave vector component $k_j$. The spin textures at a given point in reciprocal space represent the response to an effective magnetic field whose components are proportional to $\alpha_{ij} k_j$. After determination of the Bloch functions in the basis Eq. M5, spin polarization is computed as,

$$\langle \sigma_i \rangle_+^{v,c} = \frac{\langle \psi_+^{v,c} | \sigma_i | \psi_+^{v,c} \rangle}{\langle \psi_+^{v,c} | \psi_+^{v,c} \rangle}; \quad \langle \sigma_i \rangle_-^{v,c} = \frac{\langle \psi_-^{v,c} | \sigma_i | \psi_-^{v,c} \rangle}{\langle \psi_-^{v,c} | \psi_-^{v,c} \rangle}. \tag{M7}$$

By fitting the in-plane CB spin textures (the k-dependent spin polarization) computed using DFT-PBE+SOC to the analytical theory, we simultaneously determine the conduction band spin-splitting coefficients $\beta_{ij}^c$, the crystal fields, $\delta, \zeta$, and the polarization potential, $\Delta_Y$. These fits are subject to the constraint that the Γ point bandgap shift between polar *S*-NPB and non-polar rac-NPB match the value 245 meV calculated by DFT-HSE+SOC.[4] Subsequently, the valence band spin-splitting coefficients $\alpha_{ij}^v$ are determined by fitting the *ab initio* in-plane spin texture with all other parameters determined from the CB fits. The resulting parameters are given in **Supplementary Tables S-7** and **Extended Table- 2** while comparison of fitted results to *ab initio* spin polarization is given in **Supplementary Table S-8**.

**Exciton fine structure**

The fine structure of the exciton is determined by spin-dependent interactions between its constituent electron and hole. These interactions can be broken down into the short-range (SR) and long range (LR) electron-hole exchange interactions and the effective exchange interaction due to Rashba-type spin-splitting[15,16]:

$$H_{FS} = H_{SR} + H_{LR} + H_{R,ex}^{rel}. \tag{M8}$$

In centrosymmetric semiconductors the SR exchange can be written in as an effective spin interaction involving a single exchange constant, $w$:[41]

$$\widehat{H}_{spin} = \frac{w}{2}[\boldsymbol{I} - \boldsymbol{\sigma}_e \cdot \boldsymbol{\sigma}_h], \tag{M9}$$

where $\boldsymbol{I}$, $\boldsymbol{\sigma}_e$ and $\boldsymbol{\sigma}_h$ are respectively the identity operator and Pauli operators acting on the spin components of the electron and the hole Bloch functions. Here, we consider the electron/hole exchange interaction for systems where the valence and conduction bands are of mixed parity, Eq. M6. In this case, care must be taken as additional terms with distinct exchange constants appear relative to the analysis for centrosymmetric systems. In **Supplementary Sec 3.2** we apply the methodology developed by Pikus and Bir[42] and find that the SR corrections can still be expressed in terms of effective spin operators such as in Eq. M9 but several additional exchange constants emerge representing the effects of same-parity (even/even, odd/odd) and cross-parity terms. These are given in full in Eq. S3.13. For simplicity, in all of our calculations, we assume that all distinct exchange constants to be equal, in which case Eq. M9 applies. The LR exchange corrections vanish for 2D exciton states whose electric-dipole transition matrix elements lie in the plane of the inorganic layers (X, Y sublevels) but is non-vanishing for the Z-polarized 2D exciton sub-level, with correction $\Delta_Z^{LR}$ given in SI units by[23,24]

$$\Delta_Z^{LR} = \frac{e^2\hbar^2}{2\epsilon_{in}\epsilon_0\, m_0\, \hbar\omega_0} \frac{f_Z^{exc}}{S} \frac{3}{2d}. \qquad (M10)$$

Here, $f_Z^{exc}/S$ is the oscillator strength per unit area for the Z exciton, $d$ is the inorganic layer thickness, $\hbar\omega_0$ is the exciton transition energy, $\epsilon_{in}$ is the relative dielectric constant inside the inorganic layers, while $e$ is the unit charge, $m_0$ is the free electron mass, and $\epsilon_0$ is the dielectric permittivity constant. Finally, linear-in-k Rashba SOC for electrons and holes has been shown to lead in second order perturbation theory to an effective exchange interaction for excitons in Rashba systems.[14,15,17] Generalization of the "Rashba exciton" theory of Refs. [14,15,17] to 2D excitons in a chiral system of point symmetry $C_2$ leads to the effective exchange interaction (**Supplementary Sec. 3.7**):

$$H_{R,ex}^{rel} = \qquad (M11)$$
$$2A_{rel}\frac{\mu}{\hbar^2}\{\alpha_{zx}^e\alpha_{zx}^h\,\tau_z^e\,\tau_z^h + \alpha_{xx}^e\alpha_{xx}^h\,\tau_x^e\,\tau_x^h + \alpha_{yy}^e\alpha_{yy}^h\,\tau_y^e\,\tau_y^h + \alpha_{zx}^e\alpha_{xx}^h\,\tau_z^e\,\tau_x^h$$
$$+ \alpha_{xx}^e\alpha_{zx}^h\tau_x^e\,\tau_z^h\}.$$

Here, $A_{rel}$ is a numerical coefficient given in Eq. S3.60, $\mu = (1/m_e + 1/m_h)^{-1}$ is exciton reduced effective mass for in-plane motion, $\tau_i^{e(h)}$ are Pauli operators expressed in the basis $w_{1,2}^c$, or $w_{1,2}^v$ for the electron or hole respectively, and $\alpha_{i,j}^{e(h)}$ are spin-orbit coupling coefficients for the electron (hole). If the point symmetry were the non-chiral point group $C_{2v}$, the only allowed non-zero coefficients would be $\alpha_{zx}^{e(h)}$, which occur in non-chiral systems and produce spin textures with spin polarization that is always orthogonal to the wave vector **k** and are therefore non-helical.[5] Coefficients with repeated indices $\alpha_{xx}^{e(h)}, \alpha_{yy}^{e(h)}$, produce spin textures with helicity, i.e., spin expectation value with components parallel to wavevector **k**. Such terms are allowed in chiral S/R-NPB by virtue of the absence of mirror symmetries; we characterize these terms as "chiral". Using Eqs. M9-M11, We express Eq. M8 in a basis of exciton states taken in the order $|D\rangle, |Y\rangle, |X\rangle, |Z\rangle$, whose transition dipoles from the crystal ground state respectively vanish ($D$) or are aligned along the symmetry directions $y, x, z$ using basis transformation Eq. S3.20. The result is,

$$H^{fss}_{DYXZ} = \begin{pmatrix} E_D & i\Delta_{DY} & 0 & 0 \\ -i\Delta_{DY} & E_Y & 0 & 0 \\ 0 & 0 & E_X & \Delta_{XZ} \\ 0 & 0 & \Delta_{XZ} & E_Z \end{pmatrix}. \quad \text{(M12)}$$

Here, $E_D$, $E_Y$, $E_X$, and $E_Z$ are the exciton fine structure corrections for excitons in a system with $C_{2v}$ symmetry, given in Eq. S3.67, while the terms $\Delta_{DY}$, $\Delta_{XZ}$ that occur only in chiral systems are given by,

$$\Delta_{DY} = -2A_{rel}\frac{\mu}{\hbar^2}(\alpha^e_{zx}\alpha^h_{xx} - \alpha^e_{xx}\alpha^h_{zx}), \quad \Delta_{XZ} = -2A_{rel}\frac{\mu}{\hbar^2}(\alpha^e_{zx}\alpha^h_{xx} + \alpha^e_{xx}\alpha^h_{zx}) \quad \text{(M13)}$$

Once exciton states are determined by diagonalization of Eq. M12, electric dipole transition dipole matrix elements are found by computing the matrix elements $\langle \psi^{ex}_{X_i}|P|G\rangle$ of the momentum operator, $P$, between a given exciton state, $|\psi^{ex}_{X_i}\rangle$ and the crystal ground state $|G\rangle$. The magnetic dipole transition matrix elements are derived similarly by computing $\langle \psi^{ex}_{X_i}|L + 2S|G\rangle$, where $L$, $S$ are the orbital and spin angular momentum operators respectively. These matrix elements can be expressed in the form, $P_n = \langle n|P|G\rangle = i|P_K|\mathcal{K}\,\tilde{p}_n$, $M_n = \langle n|L + 2S|G\rangle = \hbar\mathcal{K}\,\tilde{m}_n$, where $\tilde{p}_n$ and $\tilde{m}_n$ are dimensionless, $P_K$ denotes the Kane momentum matrix element[35] and $\mathcal{K}$ denotes an envelope function overlap factor common to all fine structure levels. Closed form expressions for these quantities are given in **Extended Table- 3**, written in terms of the dimensionless transition matrix elements of the uncoupled basis $|D\rangle, |Y\rangle, |X\rangle, |Z\rangle$. These are, for the electric and magnetic transition dipoles,

$$\tilde{p}_D = 0, \qquad\qquad \tilde{m}_D = \sqrt{2}\,\delta_y(\Upsilon^c_{y,0} + \chi^c_{y,0} - 1)/N_w\,\hat{y},$$
$$\tilde{p}_Y = \sqrt{2}\,C_y(1 + \delta_y^2)/N_w\,\hat{y}, \qquad\qquad \tilde{m}_Y = 0,$$
$$\tilde{p}_X = \sqrt{2}\,C_x(1 - \delta_y^2)/N_w\,\hat{x}, \qquad\qquad \tilde{m}_X = i\sqrt{2}\,\delta_y(\Upsilon^c_{z,0} + \chi^c_{z,0} + 1)/N_w\,\hat{z},$$
$$\tilde{p}_Z = \sqrt{2}\,C_z(1 - \delta_y^2)/N_w\,\hat{z}, \qquad\qquad \tilde{m}_Z = -i\sqrt{2}\,\delta_y(\Upsilon^c_{x,0} + \chi^c_{x,0} + 1)/N_w\,\hat{x}, \quad \text{(M14)}$$

where $N_w$ is the normalization factor in Eq. M5, and terms $\Upsilon_{i,0}^C = 2\,C_j C_k$ and $\chi_{i,0}^C = 2C_i^2 - 1$ are given in terms of the coefficients $C_i$ in Eq. M3. Numerical evaluation of the exciton fine structure energies, transition dipoles, rotary strengths and dissymmetry for light incident normal to the plane of the inorganic layers are given for *S*-NPB and *R*-NPB in **Extended Table- 5** calculated using the material parameters listed in Extended Table- 2. The effect of including or not including distinct SOC coefficients on the exciton fine structure of *S*-NPB is shown in **Supplementary Table S-11**.

**Calculation of absorbance and CD**

The relative dielectric tensor response of the exciton for light propagating along direction $\hat{k}$, including both electric and magnetic dipole contributions, can be written as (see **Supplementary Sec 4**),

$$\epsilon(E) = \epsilon_\infty \left\{ 1 + \Delta_{LT} \sum_n \mathcal{L}(E, E_n)\ \left( \widetilde{\boldsymbol{p}}_n^{\,*} \otimes \widetilde{\boldsymbol{p}}_n \right. \right. \tag{M15}$$
$$\left. \left. + h_m\, \widetilde{\boldsymbol{p}}_n^{\,*} \otimes (\widetilde{\boldsymbol{m}}_n \times \hat{k}) + h_m (\widetilde{\boldsymbol{m}}_n \times \hat{k})^* \otimes \widetilde{\boldsymbol{p}}_n \right) \right\}.$$

Here, $\epsilon_\infty$ is the background non-resonant contribution, $\Delta_{LT}$ is the exciton longitudinal-transverse splitting parameter defined in Eq. S3.54, while the sum is taken over the set of fine structure levels, $n$, with energy $E_n$ and dimensionless electric and magnetic transition dipoles $\widetilde{\boldsymbol{p}}_n$ and $\widetilde{\boldsymbol{m}}_n$. Each term has complex Lorentzian response, $\mathcal{L}(E, E_n)$, given in Eq. S3.56, centered at $E_n$; $\otimes$ denotes the Kronecker product, and the term $h_m$ represents the ratio of the strength of the magnetic to the electric dipole light interactions. It is given by,

$$h_m \equiv \frac{\hbar k}{2|P_K|}, \tag{M16}$$

where k is the norm of the wave vector of the light wave. To analyze transverse wave propagation along the $z$ direction normal to the layers, we write the wave equation in terms of the transverse components of the electric field resulting in a reduced dielectric tensor[43,44] from which we determine the normal modes, which are, in general, elliptically polarized. While the dielectric tensor is neither Hermitian nor symmetric, we are nevertheless able to perform a normal mode decomposition for incident light of arbitrary polarization by exploiting the fact that the propagating eigenmodes are *biorthogonal* with the set of adjoint eigenmodes, which are eigenmodes of the Hermitian adjoint of the wave equation.[45,46] (See **Supplementary Sec 4.1**). Then the decadic absorbance $\mathbb{A}$ for incident light of a given polarization $\hat{e}_{inc}$ is computed for a sample of thickness $L$ as,

$$\mathbb{A}(\hat{e}_{inc}) = -Log_{10}\left[\left|\frac{E_{out}(L)}{E_{inc}}\right|^2\right]. \quad (M17)$$

The CD is then computed as the difference in decadic absorbance of incident light of left- and right- circular polarization. Analytical results for the relative rotary strengths (Eq. S3.71) of each fine structure level for light incident normal to the plane of the inorganic layers are given for chiral NPB in **Extended** Table- 4. Analytical results for the dissymmetry, $g_{CD,\perp}$, (supplementary Eq. S4.15) for illumination by light incident normal to the inorganic layers shows that this quantity is proportional to $h_m$, Eq. M16; for the parameters in **Extended Table- 2**, we find $g_{CD,\perp}$, less than 1%.

**Continuum interband transitions**

The approach we describe for computing excitonic absorbance and CD can be applied to model the optical properties associated with the continuum interband (IB) transitions by modification of Eq. M15. In **Supplementary Sec. 5**, we show that the IB dielectric response can be written as a sum over interband transitions, including electric and magnetic dipole contributions,

$$\epsilon_{IB,ed}(E) = \epsilon_\infty \Bigg\{ 1$$

$$+ \frac{e^2 \hbar^2}{2\epsilon_\infty \epsilon_0 \, m_0 \, E} \left(\frac{E_p}{E}\right) \frac{1}{\Lambda S} \sum_{\mathbf{k}} \sum_{m,n} \mathcal{L}(E, E_n) \left( \tilde{\mathbf{p}}_{\mathbf{k},m,n}^* \otimes \tilde{\mathbf{p}}_{\mathbf{k},m,n} \right. \quad (M18)$$

$$+ h_m \, \tilde{\mathbf{p}}_{\mathbf{k},m,n}^* \otimes (\tilde{\mathbf{m}}_{\mathbf{k},m,n} \times \hat{k}_L) + h_m \left( \tilde{\mathbf{m}}_{\mathbf{k},m,n} \times \hat{k}_L \right)^* \otimes \tilde{\mathbf{p}}_{\mathbf{k},m,n} \Bigg) \Bigg\}.$$

The summation is performed over all transitions from crystal ground state $|G\rangle$ resulting in creation of electron and hole pairs in state $|\varphi_{e,m,\mathbf{k}}; \varphi_{h,n,-\mathbf{k}}\rangle$ comprised of an electron with in-plane wave vector $\mathbf{k}$ in sub-band $m$ and a hole in sub-band $m$ with wave vector satisfying conservation of quasi-momentum. For light incident normal to the plane of the inorganic layers this requires that the hole quasi-momentum equal $-\mathbf{k}$. The terms $\Lambda, S$ are the interlayer spacing and the lateral area of the sample, respectively. The dimensionless transition moments $\tilde{\mathbf{p}}_{\mathbf{k},m,n}$, $\tilde{\mathbf{m}}_{\mathbf{k},m,n}$ can be directly related to the corresponding transition moments for the exciton. As a result, expressions can be derived for the relative CD rotary strength, given in **Extended Table- 6** which can be compared to those of the exciton. This analysis shows that CD vanishes unless the products $\alpha^e_{xx} \alpha^h_{zx}$ and/or $\alpha^h_{xx} \alpha^e_{zx}$ are non-zero. That is, CD requires that both the Rashba-like $\alpha^{e(h)}_{zx}$ and the chiral $\alpha^{e(h)}_{xx}$ SOC terms are present in both bands. This is the same condition as was found for the excitonic transitions (**Extended Table- 4** and **Fig. 5**) where non-zero rotary strength required non-zero values of $\Delta_{XZ}$ and/or $\Delta_{DY}$, proportional respectively to the sum and the difference of the products $\alpha^e_{xx} \alpha^h_{zx}$ and $\alpha^h_{xx} \alpha^e_{zx}$ (see Eq. M13). Importantly, the result shows that the phenomenon of CD and optical activity does not require electron/hole correlation or the electron/hole exchange interaction. Nor should the CD response be directly proportional to the spin-related sub-band energy splitting without consideration of the spin textures: The expressions for the rotary strength of IB transitions in **Extended Table- 6** are independent of the spin-splitting energy in either band. This can be seen easily in the limit $\alpha_{xx}^{e(h)} \ll \alpha_{zx}^{e(h)}$ where the roraty strengths are proportional to $\sim \alpha_{zx}^{e(h)} \alpha_{xx}^{h(e)} / \alpha_{zx}^{e} \alpha_{zx}^{h}$.

**Generalization to nanocrystals**

The formalism developed is straight-forward to generalize beyond 2D chiral perovskite systems. As an example, generalization to ferro-electric perovskite nanocrystals is detailed in **Supplementary Sec. 6**. In brief: The bulk ferro-electric CsPbBr$_3$ structure in Ref. [18] is non-chiral with a polar distortion in the $a, c$ plane (**Supplementary Table S-14**). To handle this, mixed parity Bloch functions for an arbitrarily directed polar distortion are given in Eq. S1.10, with attendant modifications in the electric dipole (Eq. S3.27) and magnetic dipole (Eqs. S3.34 & S3.46) transition matrix elements. In near-cuboidal shaped NCs the envelope function part of the exciton wavefunction Eq. M1 is replaced by one appropriate for three-dimensional intermediate confinement (Eq. S6.1), with resultant modifications to the exchange and transition dipole overlap functions (Eqs. S6.4 and S6.5 respectively) and the LR exchange corrections (Eq. S6.7). Note that the crystal structure of ferroelectric CsPbBr$_3$ is very close to a $\sqrt{2} \times \sqrt{2} \times 2$ modification relative to a cubic perovskite structure, with angle $\beta$=90.386° very near 90°; for simplicity, apart from its impact on the ferroelectric distortion, we neglect the deviation of $\beta$ from 90° in modelling the exciton shape and level structure. For the non-chiral ferro-electric CsPbBr$_3$ crystal structure, Rashba-type spin textures were reported [18]. The resulting effective exchange interaction for excitons due to Rashba splitting induced by a polar distortion along unit vector $\hat{n}$ is given by Eq. S6.12, a generalization of similar expressions in Refs. [14, 15], which can be parameterized in terms of the exciton Rashba energy,

$$\mathcal{E}_R = \mu \, \alpha^e \, \alpha^h / \hbar^2, \tag{M19}$$

written in terms of the electron and hole Rashba parameters and the reduced effective mass [15]. For oriented, dense NC assemblies, the chirooptical response can be computed from the fine structure directly using the dielectric tensor given in Eq. M15. The exciton dielectric tensor model developed allows effects such as "apparent CD"[27,28] to be modelled from the wave equation rather than within the phenomenological Muller matrix approach since the interplay of dispersion and loss effects are fully captured by the approach (see **Supplementary Sections 4** and **6**). For colloidal NCs, we utilize Maxwell Garnett theory applied to the rotationally averaged dielectric tensor as detailed in **Supplementary Sec. 6.3**.

**Method references**

## Data availability

All data that support the findings of this study are available from the corresponding author upon request.

## Acknowledgements

This work was authored by the Alliance for Sustainable Energy, LLC, the manager and operator of the National Renewable Energy Laboratory for DOE under contract no. DE-AC36-08GO28308, and was primarily supported through the Center for Hybrid Organic–Inorganic

Semiconductors for Energy (CHOISE), an Energy Frontier Research Center funded by the Office of Basic Energy Sciences, Office of Science within the US Department of Energy through contract number DE-AC36-08G028308. The views expressed in the article do not necessarily represent the views of the DOE or the U.S. Government. This research used resources of the National Energy Research Scientific Computing Center (NERSC), a U.S. Department of Energy Office of Science User Facility supported by the Office of Science of the U.S. Department of Energy under Contract No. DE-AC02-05CH11231 using NERSC award BES-ERCAP0029145.

## Author contributions

P.C.S. and M.C.B. conceived the project. R.S. and V.B. carried out first principles DFT calculations. M.P.H. carried out the synthesis and optical measurements on chiral NPB. P.C.S. developed the K.P and effective mass theory for polar perovskites and the wave propagation theory, analyzed the DFT band structure and spin textures, and wrote the manuscript with assistance in figure preparation from R.S and M.P.H.

## Additional information

Supplementary information is available in the online version of the paper. Reprints and permissions information is available online at www.nature.com/reprints. Correspondence and requests for materials should be addressed to P.C.S.

## Competing financial interests

The authors declare no competing financial interests.

# Extended Data Figures

**Extended Table- 1.** Analysis of formal local dipole moments in S-NPB, experimental structure
Part (a) gives the primitive vectors of the experimentally determined structure of S-NPB, space group $P2_1$, as reported in Supplementary Table 35 of Ref. [5]. Vectors $a$ and $b$ lie in the plane of the inorganic layers with $b$ parallel to the $2_1$ screw axis, defining the $y$ axis, with the $x$ axis parallel to vector $a$. Vector $c$ is the layer stacking direction and points largely out of plane, with angle $\beta$ between $a$, $c$ equal to 93.805 degrees. Part (b) shows the total local formal dipole associated with each Pb atom within a unit cell, and their resultant. Parts (c, d) give the local formal dipole analysis for each Pb site within a unit cell. The local dipole moment associated with each Pb site, set as origin, is calculated from the coordinates of the surrounding nearest six Br atoms using the general formula: $\sum q_{Br}(r_{Br} - r_{Pb})$ (here, $q_{Br}$ is taken as $-1\,e$ where e is the elementary charge). Atomic positions in the table are given in Angstroms in cartesian coordinates. The corresponding average of the local polarization within each $PbBr_6$ octahedron is 8.9 $\mu C/cm^2$ directed along $y$, parallel to the $2_1$ screw axis.

(a)

|   | x | y | z |
|---|---|---|---|
| a | 8.724 | 0 | 0 |
| b | 0 | 7.930 | 0 |
| c | -1.290 | 0 | 19.398 |

(b)

|   | B-sites | Px (e-Å) | Py (e-Å) | Pz (e-Å) |
|---|---|---|---|---|
| local dipole | Pb1 | 0.124 | 0.200 | -0.231 |
| local dipole | Pb2 | -0.124 | 0.200 | 0.231 |
| **Sum (e-Å)** | **Pb1+Pb2** | **0.000** | **0.399** | **0.000** |

(c)

|   | x | y | z | x | y | z |
|---|---|---|---|---|---|---|
| Pb1 | 1.510 | 0.296 | 9.642 | 0.000 | 0.000 | 0.000 |
| Br1 | 1.653 | 0.329 | 12.553 | 0.143 | 0.033 | 2.912 |
| Br2 | 4.050 | 1.868 | 9.443 | 2.540 | 1.572 | -0.198 |
| Br3 | 0.050 | 2.918 | 9.502 | -1.460 | 2.622 | -0.140 |
| Br4 | 1.140 | -0.397 | 6.730 | -0.370 | -0.693 | -2.911 |
| Br5 | 3.384 | -2.097 | 9.955 | 1.874 | -2.393 | 0.313 |
| Br6 | -1.340 | -1.046 | 9.897 | -2.850 | -1.342 | 0.255 |
| local dipole (e-Å) |   |   |   | 0.124 | 0.200 | -0.231 |

(d)

|   | x | y | z | x | y | z |
|---|---|---|---|---|---|---|
| Pb2 | 5.924 | 4.261 | 9.757 | 0.000 | 0.000 | 0.000 |
| Br1 | 5.781 | 4.294 | 6.845 | -0.143 | 0.033 | -2.912 |
| Br2 | 6.294 | 3.568 | 12.668 | 0.370 | -0.693 | 2.911 |
| Br3 | 3.384 | 5.833 | 9.955 | -2.540 | 1.572 | 0.198 |
| Br4 | 4.050 | 1.868 | 9.443 | -1.874 | -2.393 | -0.313 |
| Br5 | 7.384 | 6.883 | 9.897 | 1.460 | 2.622 | 0.140 |
| Br6 | 8.774 | 2.918 | 9.502 | 2.850 | -1.342 | -0.255 |
| local dipole (e-Å) |   |   |   | -0.124 | 0.200 | 0.231 |

**Extended Table- 2.** Material parameters and exciton properties calculated or assumed for S-NPB.

| Parameter | Value | Source |
|---|---|---|
| 1s exciton transition energy | 3163 meV | Measured |
| Inorganic well thickness, $d$ | $d = 0.6$ nm | Ref.[4] |
| Organic barrier layer thickness, $b$ | $b = 1.34$ nm | Ref.[4] |
| Organic layer relative permittivity $\epsilon_{org}$ | $\epsilon_{org} = 2.56$ | Ref. [47] |
| Ordinary axis refractive index, $n_o$ | $n_o = 1.91$ | Ref.[48] |
| Effective background relative permittivity $\epsilon_\infty$ | $\epsilon_\infty = 3.65$ | $\epsilon_\infty = n_o^2$ |
| Inorganic layer background relative dielectric permittivity $\epsilon_{in}$ | $\epsilon_{in} = 6.1$ | $\epsilon_{in} = [(d+b)\epsilon_{org} - b\epsilon_{in}]/d$ |
| Exciton reduced effective mass | $\mu = 0.239\, m_0$ | **Supplementary Table S-5** |
| Calculated 1s exciton binding energy, $B_{1,0}$<br>2D exciton radius, $a_{10}$ | $B_{1,0} = 520$ meV<br>$a_{10} = 1.6$ nm | This work |
| Calculated 2p exciton binding energy, $B_{2,1}$<br>2D exciton radius, $a_{21}$ | $B_{2,1} = 165$ meV<br>$a_{21} = 0.75$ nm | This work |
| SR exchange constant, $w_{SR}$ | $w_{SR} = 30$ meV | Ref.[49] |
| Kane energy, $E_p$ | $E_p = 5.5$ eV | Ref.[50] |
| Longitudinal-transverse splitting parameter $\Delta_{LT}$ | $\Delta_{LT} = 53$ meV | Suppl. Eq. S3.54 |
| Split-off parameter $\Delta$<br>"Tetragonal" Crystal field: $\delta$<br>"Orthorhombic" Crystal field: $\zeta$ | $\Delta = 1350$ meV<br>$\delta = -940.5$ meV<br>$\zeta = -92.6$ meV | **Supplementary Table S-7** |
| Polarization potential $\Delta_Y$<br>Parity mixing amplitude, $\delta_Y$ | $\Delta_Y = 929.7$ meV<br>$\delta_Y = 0.186$ | **Supplementary Table S-7**<br>Suppl. Eq. S1.12 |
| Electron spin splitting coefficients<br>$\alpha^e_{zx}, \alpha^e_{xx}, \alpha^e_{yy}$ | $\alpha^e_{zx} = -140.4$ meV·nm<br>$\alpha^e_{xx} = -45.3$ meV·nm<br>$\alpha^e_{yy} = 28.5$ meV·nm | **Supplementary Table S-7** |
| Hole spin splitting coefficients<br>$\alpha^h_{zx}, \alpha^h_{xx}, \alpha^h_{yy}$ | $\alpha^h_{zx} = 31.7$ meV·nm<br>$\alpha^h_{xx} = -1.8$ meV·nm<br>$\alpha^h_{yy} = -0.3$ meV·nm | **Supplementary Table S-7** |
| Relative motion numerical factor, $A_{rel}$ | $A_{rel} = 0.305$ | Suppl. Eq. S3.60 |
| Ratio of magnetic to electric interaction strength, $h_m$ | $h_m = 0.0033$ | Eq. M1 |

**Extended Table- 3**. Electric dipole, magnetic dipole and electric quadrupole transition matrix elements for chiral 2D exciton.

Within the tables, the set of four states $n$ comprises the two levels, $\mathcal{D}$ and $\mathcal{Y}$, corresponding to irrep $A$ in **Supplementary Table S-10**, which are admixtures of the two uncoupled $D, Y$ eigenstates in the absence of the effective exchange interaction due to spin splitting, and the two levels, $\mathcal{X}$ and $\mathcal{Z}$, corresponding to irrep $B$ in **Supplementary Table S-10** which are each admixtures of the two uncoupled $X, Z$ states.

**Panel (a):** The transition dipoles $\widetilde{\boldsymbol{p}}_n$ and $\widetilde{\boldsymbol{m}}_n$ are expressed in dimensionless form and are related to the dimensioned transition dipoles through $\boldsymbol{P}_n = i|P_K|\mathcal{K}\, \widetilde{\boldsymbol{p}}_n$, and $\boldsymbol{M}_n = \hbar\mathcal{K}\, \widetilde{\boldsymbol{m}}_n$ in terms of $P_K$, the Kane momentum matrix element and $\mathcal{K}$, the overlap factor (Eq. S3.29), common to all fine structure levels. Transition dipoles $\widetilde{\boldsymbol{p}}_n$ and $\widetilde{\boldsymbol{m}}_n$ of the coupled levels are given in terms of the dimensionless transition dipoles, $\tilde{p}_{X_i}$ and $\tilde{m}_{X_i}$, of the uncoupled states given in Eq. M14, where $X_i$ denotes the uncoupled $D, X, Y, Z$ basis. Terms $\Delta_{DY}, \Delta_{XZ}$ and $\mathcal{N}_{DY}, \mathcal{N}_{XZ}$ are given in Supplementary Eqs. M13/S3.68 and Eq.S3.70 respectively. Energies $E_i$ and $\mathcal{E}_j$ are given in Supplementary Eqs. S3.67 and S3.69, respectively. Parity mixing amplitude $\delta_Y$ is defined in Supplementary Eq. S1.12.

| $n$ | E | Electric transition dipole, $\widetilde{\boldsymbol{p}}_n$, where $\boldsymbol{P}_n = \langle n|\boldsymbol{P}|G\rangle = i|P_K|\mathcal{K}\,\widetilde{\boldsymbol{p}}_n$ | Magnetic transition dipole $\widetilde{\boldsymbol{m}}_n$ where $\boldsymbol{M}_n = \langle n|\boldsymbol{M}|G\rangle = \hbar\mathcal{K}\,\widetilde{\boldsymbol{m}}_n$ |
|---|---|---|---|
| $\mathcal{D}$ | $\mathcal{E}_\mathcal{D}$ | $\dfrac{1}{\sqrt{\mathcal{N}_{DY}}} \dfrac{\Delta_{DY}}{(E_Y - \mathcal{E}_\mathcal{D})} (\tilde{p}_Y \cdot \hat{y})\, \hat{y}$ | $\dfrac{i\,\delta_Y}{\sqrt{\mathcal{N}_{DY}}} \mathrm{Re}\left(\dfrac{\widetilde{m}_D \cdot \hat{y}}{|\delta_Y|}\right) \hat{y}$ |
| $\mathcal{Y}$ | $\mathcal{E}_\mathcal{Y}$ | $\dfrac{1}{\sqrt{\mathcal{N}_{DY}}} (\tilde{p}_Y \cdot \hat{y})\, \hat{y}$ | $\dfrac{-i\,\delta_Y}{\sqrt{\mathcal{N}_{DY}}} \mathrm{Re}\left(\dfrac{\Delta_{DY}}{(E_Y - \mathcal{E}_\mathcal{D})} \dfrac{\widetilde{m}_D \cdot \hat{y}}{|\delta_Y|}\right)\hat{y}$ |
| $\mathcal{X}$ | $\mathcal{E}_\mathcal{X}$ | $\dfrac{1}{\sqrt{\mathcal{N}_{XZ}}}\left( (\tilde{p}_X \cdot \hat{x})\hat{x} - \dfrac{\Delta_{XZ}}{(E_Z - \mathcal{E}_\mathcal{X})}(\tilde{p}_Z \cdot \hat{z})\hat{z} \right)$ | $\dfrac{i\,\delta_Y}{\sqrt{\mathcal{N}_{XZ}}} \mathrm{Im}\left( \dfrac{-\Delta_{XZ}}{(E_Z - \mathcal{E}_\mathcal{X})} \dfrac{\widetilde{m}_Z \cdot \hat{x}}{|\delta_Y|}\hat{x} + \dfrac{\widetilde{m}_X \cdot \hat{z}}{|\delta_Y|}\hat{z} \right)$ |
| $\mathcal{Z}$ | $\mathcal{E}_\mathcal{Z}$ | $\dfrac{1}{\sqrt{\mathcal{N}_{XZ}}}\left( \dfrac{\Delta_{XZ}}{(E_Z - \mathcal{E}_\mathcal{X})}(\tilde{p}_X \cdot \hat{x})\hat{x} + (\tilde{p}_Z \cdot \hat{z})\hat{z} \right)$ | $\dfrac{i\,\delta_Y}{\sqrt{\mathcal{N}_{XZ}}} \mathrm{Im}\left( \dfrac{\widetilde{m}_Z \cdot \hat{x}}{|\delta_Y|}\hat{x} + \dfrac{\Delta_{XZ}}{(E_Z - \mathcal{E}_\mathcal{X})} \dfrac{\widetilde{m}_X \cdot \hat{z}}{|\delta_Y|}\hat{z} \right)$ |

**Panel (b):** The electric quadrupole transition matrix elements $\widetilde{\boldsymbol{Q}}_n$ are expressed in dimensionless form and are related to the dimensioned $\boldsymbol{Q}_n = \hbar\mathcal{K}\, \widetilde{\boldsymbol{Q}}_n$ in terms of $\mathcal{K}$, the overlap factor, common to all fine structure levels. Matrix elements given in terms of the components $\tilde{Q}_{i,j}$, of the uncoupled states $X_i$ listed in **Supplementary Table S-9**, where $X_i$ denotes the uncoupled $D, X, Y, Z$ basis in the absence of spin splitting terms.

| $n$ | E | Electric quadrupole transition matrix elements $\widetilde{Q}_{n;i,j}$ where $\boldsymbol{Q}_n = \langle n|\boldsymbol{Q}|G\rangle = \hbar\mathcal{K}\, \widetilde{\boldsymbol{Q}}_n$ |
|---|---|---|
| $\mathcal{D}$ | $\mathcal{E}_\mathcal{D}$ | $\dfrac{i\,\delta_Y}{|\delta_Y|\sqrt{\mathcal{N}_{DY}}} Q_{xz};\ \dfrac{\delta_Y}{|\delta_Y|\sqrt{\mathcal{N}_{DY}}} \dfrac{\Delta_{DY}}{(E_Y - \mathcal{E}_\mathcal{D})} Q_{xx};\ \dfrac{\delta_Y}{|\delta_Y|\sqrt{\mathcal{N}_{DY}}} \dfrac{\Delta_{DY}}{(E_Y - \mathcal{E}_\mathcal{D})} Q_{yy};\ \dfrac{\delta_Y}{|\delta_Y|\sqrt{\mathcal{N}_{DY}}} \dfrac{\Delta_{DY}}{(E_Y - \mathcal{E}_\mathcal{D})} Q_{z,z}$ |
| $\mathcal{Y}$ | $\mathcal{E}_\mathcal{Y}$ | $\dfrac{-i\,\delta_Y}{|\delta_Y|\sqrt{\mathcal{N}_{DY}}} \dfrac{\Delta_{DY}}{(E_Y - \mathcal{E}_\mathcal{D})} Q_{xz};\ \dfrac{\delta_Y}{|\delta_Y|\sqrt{\mathcal{N}_{DY}}} Q_{xx};\ \dfrac{\delta_Y}{|\delta_Y|\sqrt{\mathcal{N}_{DY}}} Q_{yy};\ \dfrac{\delta_Y}{|\delta_Y|\sqrt{\mathcal{N}_{DY}}} Q_{zz}$ |
| $\mathcal{X}$ | $\mathcal{E}_\mathcal{X}$ | $\dfrac{\delta_Y}{|\delta_Y|\sqrt{\mathcal{N}_{XZ}}} \left( \dfrac{-\Delta_{XZ}}{(E_Z - \mathcal{E}_\mathcal{X})} Q_{yz} + Q_{xy} \right)$ |
| $\mathcal{Z}$ | $\mathcal{E}_\mathcal{Z}$ | $\dfrac{\delta_Y}{|\delta_Y|\sqrt{\mathcal{N}_{XZ}}} \left( Q_{yz} + \dfrac{\Delta_{XZ}}{(E_Z - \mathcal{E}_\mathcal{X})} Q_{xy} \right)$ |

**Extended Table- 4**. Rotary strength of CD at normal incidence for chiral 2D exciton

The table shows the relative rotary strength of the CD at normal incidence for each of the set of four fine structure states, $n$, comprising the two coupled $\mathcal{D}$ and $\mathcal{Y}$ levels, which correspond to irrep $A$, and the two coupled $\mathcal{X}, \mathcal{Z}$ levels, irrep $B$, referring to the irreducible representation labels of **Supplementary Table S-10**.

**Panel (a):** Relative strength of CD at normal incidence, $\tilde{\mathcal{R}}_{CD,\perp}$, associated with coupled electric dipole/magnetic dipole transitions. Expressions are given in terms of the dimensionless transition dipoles, $\tilde{p}_{X_i}$ and $\tilde{m}_{X_i}$, of the uncoupled states given in Eq. M14, where $X_i$ denotes the uncoupled $D, X, Y, Z$ basis; the parity mixing amplitude $\delta_Y$, given in Supplementary Eq. S1.12, and the terms $\Delta_{DY}, \Delta_{XZ}$ and $\mathcal{N}_{DY}, \mathcal{N}_{XZ}$ given in Supplementary Eqs. S3.68 and S3.70 respectively. Energies $E_i$ and $\mathcal{E}_j$ are given in Supplementary Eqs. S3.67 and S3.69, respectively.

| State $n$ | Energy Eq. S3.61 | $\tilde{\mathcal{R}}_{CD,\perp}$ (ED/MD) $= \text{Im}\left(\tilde{p}_{n,x}\tilde{m}_{n,x} + \tilde{p}_{n,y}\tilde{m}_{n,y}\right)$ |
|---|---|---|
| $\mathcal{D}$ | $\mathcal{E}_\mathcal{D}$ | $\dfrac{1}{\mathcal{N}_{DY}}\left(\dfrac{\Delta_{DY}\,\delta_Y}{(E_Y - \mathcal{E}_\mathcal{D})}\,(\tilde{p}_Y \cdot \hat{y})\,\dfrac{\text{Re}(\tilde{m}_D \cdot \hat{y})}{|\delta_Y|}\right)$ |
| $\mathcal{Y}$ | $\mathcal{E}_\mathcal{Y}$ | $-\dfrac{1}{\mathcal{N}_{DY}}\left(\dfrac{\Delta_{DY}\,\delta_Y}{(E_Y - \mathcal{E}_\mathcal{D})}\,(\tilde{p}_Y \cdot \hat{y})\,\dfrac{\text{Re}(\tilde{m}_D \cdot \hat{y})}{|\delta_Y|}\right)$ |
| $\mathcal{X}$ | $\mathcal{E}_\mathcal{X}$ | $-\dfrac{1}{\mathcal{N}_{XZ}}\left(\dfrac{\Delta_{XZ}\,\delta_Y}{(E_Z - \mathcal{E}_\mathcal{X})}\,(\tilde{p}_X \cdot \hat{x})\,\dfrac{\text{Im}(\tilde{m}_Z \cdot \hat{x})}{|\delta_Y|}\right)$ |
| $\mathcal{Z}$ | $\mathcal{E}_\mathcal{Z}$ | $\dfrac{1}{\mathcal{N}_{XZ}}\left(\dfrac{\Delta_{XZ}\,\delta_Y}{(E_Z - \mathcal{E}_\mathcal{X})}\,(\tilde{p}_X \cdot \hat{x})\,\dfrac{\text{Im}(\tilde{m}_Z \cdot \hat{x})}{|\delta_Y|}\right)$ |

**Panel (b):** Relative rotary strength of CD at normal incidence associated with coupled electric dipole/electric quadrupole transitions. Expressions are given in terms of the dimensionless electric transition dipoles, $\tilde{p}_{X_i}$, of the uncoupled states given in Eq. M14 and the dimensionless quadrupole matrix elements $\tilde{Q}_{X_i}$, from **Supplementary Table S-9**, where $X_i$ denotes the uncoupled $D, X, Y, Z$ basis; the parity mixing amplitude $\delta_Y$, given in Supplementary Eq. S1.12, and the terms $\Delta_{DY}, \Delta_{XZ}$ and $\mathcal{N}_{DY}, \mathcal{N}_{XZ}$ given in Supplementary Eqs. S3.68 and S3.70 respectively. Energies $E_i$ and $\mathcal{E}_j$ are given in Supplementary Eqs. S3.67 and S3.69, respectively.

| State $n$ | Energy (Eq. S3.61) | $\tilde{\mathcal{R}}_{CD,\perp}$ (ED/QE) $= \text{Im}\left(\tilde{p}_{n,y}\tilde{Q}_{n;x,z} - \tilde{p}_{n,x}\tilde{Q}_{n;y,z}\right)$ |
|---|---|---|
| $\mathcal{D}$ | $\mathcal{E}_\mathcal{D}$ | $\dfrac{1}{\mathcal{N}_{DY}}\left(\dfrac{\Delta_{DY}\,\delta_Y}{(E_Y - \mathcal{E}_\mathcal{D})}\,(\tilde{p}_Y \cdot \hat{y})\,\dfrac{\text{Re}\,Q_{x,z}}{|\delta_Y|}\right)$ |
| $\mathcal{Y}$ | $\mathcal{E}_\mathcal{Y}$ | $-\dfrac{1}{\mathcal{N}_{DY}}\left(\dfrac{\Delta_{DY}\,\delta_Y}{(E_Y - \mathcal{E}_\mathcal{D})}\,(\tilde{p}_Y \cdot \hat{y})\,\dfrac{\text{Re}\,Q_{x,z}}{|\delta_Y|}\right)$ |
| $\mathcal{X}$ | $\mathcal{E}_\mathcal{X}$ | $\dfrac{1}{\mathcal{N}_{XZ}}\left(\dfrac{\Delta_{XZ}\,\delta_Y}{(E_Z - \mathcal{E}_\mathcal{X})}\,(\tilde{p}_X \cdot \hat{x})\,\dfrac{\text{Im}\,Q_{y,z}}{|\delta_Y|}\right)$ |
| $\mathcal{Z}$ | $\mathcal{E}_\mathcal{Z}$ | $-\dfrac{1}{\mathcal{N}_{XZ}}\left(\dfrac{\Delta_{XZ}\,\delta_Y}{(E_Z - \mathcal{E}_\mathcal{X})}\,(\tilde{p}_X \cdot \hat{x})\,\dfrac{\text{Im}\,Q_{y,z}}{|\delta_Y|}\right)$ |

**Extended Table- 5**. Comparison of fine structure in S- and R-NPB

The table compares the exciton fine structure level energies and electric and magnetic transition dipole moments in *S*-NPB, panel (a), and *R*-NPB, panel (b). Calculations include SR and LR exchange plus the effective exchange interaction due to Rashba-like spin-splitting for in-plane dispersion including all terms found by analysis of the spin textures calculated in DFT-PBE+SOC. The transition dipoles $\tilde{\boldsymbol{p}}_n$ and $\tilde{\boldsymbol{m}}_n$ are expressed in dimensionless form and are related to the dimensioned transition dipoles through $\boldsymbol{P}_n = i|P_K|\mathcal{K}\,\tilde{\boldsymbol{p}}_n$, and $\boldsymbol{M}_n = \hbar\mathcal{K}\,\tilde{\boldsymbol{m}}_n$ in terms of $P_K$, the Kane momentum matrix element and $\mathcal{K}$, the overlap factor, common to all fine structure levels. The relative magnitude, $\tilde{\mathcal{R}}_{CD,\perp}$, of the CD at normal incidence for each of the set of four fine structure states, $n$, given by $\tilde{\mathcal{R}}_{CD,\perp} = \mathrm{Im}(\tilde{p}_{n,x}\tilde{m}_{n,x} + \tilde{p}_{n,y}\tilde{m}_{n,y})$, Supplementary Eq. S3.71, is shown in the table. The maximum dissymmetry, $g_{CD}$, in the limit of vanishing line width, given by Supplementary Eq. S4.15, is given as well. Panel (a) shows fine structure calculated for *S*-NPB with all terms using parameters in **Extended Table- 2**. Panel (b) shows fine structure calculated for *R*-NPB, using the same parameters but reversing the sign of the parity mixing potential $\Delta_Y$, the parity mixing amplitude, $\delta_Y$, and the signs of all SOC coefficients $\alpha_{xx}^{e,h}, \alpha_{yy}^{e,h}, \alpha_{zx}^{e,h}$ relative to the values for S-NPB. The sign reversal of $\Delta_Y, \delta_Y$ is due to the reversal of the direction of the polar distortion (**Fig. 1**) and causes a sign change in the magnetic transition dipoles in *R*-NPB relative to *S*-NPB. The sign change in the SOC coefficients reflects the reversal of the spin textures in *R*-NPB relative to *S*-NPB, as shown in **Fig 1**.

**(a) S-NPB**

| State | E (meV) | $\tilde{p}_x$ | $\tilde{p}_y$ | $\tilde{p}_z$ | $\mathrm{Im}(\tilde{m}_x)$ | $\mathrm{Im}(\tilde{m}_y)$ | $\mathrm{Im}(\tilde{m}_z)$ | $\tilde{\mathcal{R}}_{CD,\perp}$ | $g_{CD} \times 10^3$ |
|---|---|---|---|---|---|---|---|---|---|
| $\mathcal{Z}$ | 51.74 | 0.047 | 0 | 0.476 | -0.315 | 0 | 0.017 | -0.015 | -88.27 |
| $\mathcal{Y}$ | 22.61 | 0 | 0.979 | 0 | 0 | -0.031 | 0 | -0.030 | -0.42 |
| $\mathcal{X}$ | 13.12 | 0.797 | 0 | -0.028 | 0.019 | 0 | 0.283 | 0.015 | 0.31 |
| $\mathcal{D}$ | 7.64 | 0 | -0.221 | 0 | 0 | -0.137 | 0 | 0.030 | 8.16 |

**(b) R-NPB**

| State | E (meV) | $\tilde{p}_x$ | $\tilde{p}_y$ | $\tilde{p}_z$ | $\mathrm{Im}(\tilde{m}_x)$ | $\mathrm{Im}(\tilde{m}_y)$ | $\mathrm{Im}(\tilde{m}_z)$ | $\tilde{\mathcal{R}}_{CD,\perp}$ | $g_{CD} \times 10^3$ |
|---|---|---|---|---|---|---|---|---|---|
| $\mathcal{Z}$ | 51.74 | 0.047 | 0 | 0.476 | 0.315 | 0 | -0.017 | 0.015 | 88.27 |
| $\mathcal{Y}$ | 22.61 | 0 | 0.979 | 0 | 0 | 0.031 | 0 | 0.030 | 0.42 |
| $\mathcal{X}$ | 13.12 | 0.797 | 0 | -0.028 | -0.019 | 0 | -0.283 | -0.015 | -0.31 |
| $\mathcal{D}$ | 7.64 | 0 | -0.221 | 0 | 0 | 0.137 | 0 | -0.030 | -8.16 |

**Extended Table- 6**. Rotary strength for IB continuum transitions in chiral 2D semiconductor: Coupled ED/MD transitions.

The table shows the transition energy and relative rotary strength of the CD at normal incidence, $\tilde{\mathcal{R}}_{\perp;\boldsymbol{k},m,n}$, associated with coupled electric dipole/magnetic dipole transitions for each of the four interband continuum absorption transitions from the crystal ground state, state $|G\rangle$ to uncorrelated electron and hole pair state $|\varphi_{e,m,k_x};\varphi_{h,n,-k_x}\rangle$. These pair states are comprised of an electron with in-plane wave vector $\boldsymbol{k} = k_x \hat{x}$ in sub-band $m$ and a hole with $\boldsymbol{k} = -k_x \hat{x}$ in sub-band $m$. Transition energies are expressed in terms of the electron and hole energies given in Supplementary Eq. S5.4. The terms $\alpha_{eff}^{e(h)} = \sqrt{\left(\alpha_{xx}^{e(h)}\right)^2 + \left(\alpha_{zx}^{e(h)}\right)^2}$. Expressions are given in terms of the dimensionless transition dipoles, $\tilde{p}_{X_i}$ and $\tilde{m}_{X_i}$, given in Eq. M14, corresponding to zone-center transition moments expressed in the $D, X, Y, Z$ basis, denoted $X_i$. The term $\delta_Y$ denotes the parity mixing amplitude given in Supplementary Eq. S1.12.

| Transition energy | $\tilde{\mathcal{R}}_{\perp;\boldsymbol{k},m,n}(\text{ED/MD}) = \text{Im}\left[\left(\tilde{\boldsymbol{p}}_{\boldsymbol{k},m,n}\right)_x^* \left(\tilde{\boldsymbol{m}}_{\boldsymbol{k},m,n}\right)_x + \left(\tilde{\boldsymbol{p}}_{\boldsymbol{k},m,n}\right)_y^* \left(\tilde{\boldsymbol{m}}_{\boldsymbol{k},m,n}\right)_y\right]$ |
|---|---|
| $E_{1,k_x}^e + E_{1,-k_x}^h$ | $+\dfrac{\delta_Y}{\alpha_{eff}^e \alpha_{eff}^h}\left\{\left((\alpha_{zx}^e \alpha_{xx}^h - \alpha_{xx}^e \alpha_{zx}^h)(\tilde{p}_Y \cdot \hat{y})\dfrac{\text{Re}(\tilde{m}_D \cdot \hat{y})}{|\delta_Y|}\right) - \left((\alpha_{zx}^e \alpha_{xx}^h + \alpha_{xx}^e \alpha_{zx}^h)(\tilde{p}_X \cdot \hat{x})\dfrac{\text{Im}(\tilde{m}_Z \cdot \hat{x})}{|\delta_Y|}\right)\right\}$ |
| $E_{1,k_x}^e + E_{2,-k_x}^h$ | $-\dfrac{\delta_Y}{\alpha_{eff}^e \alpha_{eff}^h}\left\{\left((\alpha_{zx}^e \alpha_{xx}^h - \alpha_{xx}^e \alpha_{zx}^h)(\tilde{p}_Y \cdot \hat{y})\dfrac{\text{Re}(\tilde{m}_D \cdot \hat{y})}{|\delta_Y|}\right) - \left((\alpha_{zx}^e \alpha_{xx}^h + \alpha_{xx}^e \alpha_{zx}^h)(\tilde{p}_X \cdot \hat{x})\dfrac{\text{Im}(\tilde{m}_Z \cdot \hat{x})}{|\delta_Y|}\right)\right\}$ |
| $E_{2,k_x}^e + E_{1,-k_x}^h$ | $-\dfrac{\delta_Y}{\alpha_{eff}^e \alpha_{eff}^h}\left\{\left((\alpha_{zx}^e \alpha_{xx}^h - \alpha_{xx}^e \alpha_{zx}^h)(\tilde{p}_Y \cdot \hat{y})\dfrac{\text{Re}(\tilde{m}_D \cdot \hat{y})}{|\delta_Y|}\right) - \left((\alpha_{zx}^e \alpha_{xx}^h + \alpha_{xx}^e \alpha_{zx}^h)(\tilde{p}_X \cdot \hat{x})\dfrac{\text{Im}(\tilde{m}_Z \cdot \hat{x})}{|\delta_Y|}\right)\right\}$ |
| $E_{2,k_x}^e + E_{2,-k_x}^h$ | $+\dfrac{\delta_Y}{\alpha_{eff}^e \alpha_{eff}^h}\left\{\left((\alpha_{zx}^e \alpha_{xx}^h - \alpha_{xx}^e \alpha_{zx}^h)(\tilde{p}_Y \cdot \hat{y})\dfrac{\text{Re}(\tilde{m}_D \cdot \hat{y})}{|\delta_Y|}\right) - \left((\alpha_{zx}^e \alpha_{xx}^h + \alpha_{xx}^e \alpha_{zx}^h)(\tilde{p}_X \cdot \hat{x})\dfrac{\text{Im}(\tilde{m}_Z \cdot \hat{x})}{|\delta_Y|}\right)\right\}$ |

**Extended Table- 7**. Material parameters and exciton properties for ferroelectric CsPbBr3.

Exciton and material parameters for polar phase CsPbBr$_3$ from Ref. [18]. Where parameters are unknown for the polar phase we have taken values for the orthorhombic phase computed in Ref. [15] or determined experimentally in Ref. [51]. The term $a_o$ is the hydrogen Bohr radius.

| Parameter | Value | Source |
|---|---|---|
| Polar phase band gap, $E_g$ | 2680 meV | Ref. 18 |
| Non-polar phase band gap | 2250 meV | Ref. 18 |
| Polarization potential $\Delta_X$ | -1172 meV | **Supplementary Table S-13** & This work |
| Polarization potential $\Delta_Y$ | 0 | |
| Polarization potential $\Delta_Z$ | 543 meV | |
| Exciton reduced effective mass $\mu/m_0$ | 0.177 | Ref. 18 |
| Spin-orbit split-off parameter, $\Delta$ | 1543 meV | Ref. 15 |
| Tetragonal crystal field, $\delta$ | -5.5 meV | Ref. 15 |
| Orthorhombic crystal field, $\xi$ | 45.5 meV | Ref. 15 |
| Kane energy, $E_p$ | 15 eV | Ref. 51 |
| Exciton effective dielectric constant $\epsilon_{eff}$ | 8.7 | Ref. 51 |
| Exciton Bohr radius, $a_x$ | 2.59 nm | $a_x = a_o \epsilon_{eff}/\mu$ |
| Exchange constant, | 108.9 meV | Ref. 15 |
| Bulk LT splitting, $\hbar\omega_{LT}$ | 5.3 meV | Ref. 51 |
| Electron Rashba coefficient $\alpha_R^e$ | 11.3 meV·nm | Ref. 18 |
| Hole Rashba coefficient $\alpha_R^h$ | 51.5 meV·nm | Ref. 18 |
| LT splitting parameter, $\Delta_{LT,NC}$, 10nm edge length | 2.23 meV | Supplementary Eq. S6.6 |
| High frequency dielectric constant $\epsilon_{NC,\infty}$ | 4.3 | Ref. 51 |
| Medium dielectric $\epsilon_{med} = n^2$ (polystyrene) | 2.56 | Ref. 52 |
| Ratio of magnetic to electric interaction strength, $h_m$ | $h_m = 0.0014$ | Eq. M1, Suppl. Eq. S4.11 |

**Extended Table- 8**. Exciton fine structure and transition dipoles- ferroelectric CsPbBr3 NCs

The table shows the energies and cartesian components of the electric dipole transition vectors for ferroelectric CsPbBr$_3$ NCs with $L_e$=10nm and different basal edge length ratios $q = L_y/L_x$, calculated using the parameters of Extended **Table- 7** and including the Rashba effect. Coordinates are defined in reference to the NC edges, see schematic in **Fig. 6** panel (c). The set of states are labelled according the major component of the exciton transition dipole. Components of the dimensionless transition dipoles $\tilde{p}_{X_i}$ and $\tilde{m}_{X_i}$ are given, where the full transition dipoles are given by $P_{X_i} = i|P_K|\mathcal{K}\tilde{p}_{X_i}$, and $M_{X_i} = \hbar\mathcal{K}\tilde{m}_{X_i}$ in terms of $P_K$, the Kane momentum matrix element and $\mathcal{K}$, the overlap factor (Supplementary Eq. S6.5), common to all fine structure levels. The maximum possible orientationally averaged dissymmetry, proportional to the inner product $\tilde{p}.\tilde{m}$, is also given for each transition (see Supplementary Eq. S6.20); these are of order $10^{-5}$. The angles in degrees (deg) between the projections into the $x,y$ plane of the electric dipole transition vectors are given for all pairs of states for which electric dipole transition matrix elements are non-zero.

(a) q = 1.2

| State | E (meV) | $\tilde{p}_x$ | $\tilde{p}_y$ | $\tilde{p}_z$ | $\tilde{m}_x$ | $\tilde{m}_y$ | $\tilde{m}_z$ | $\langle g_{CD,n}\rangle$ x10$^5$ | | X | Z | Y | D |
|---|---|---|---|---|---|---|---|---|---|---|---|---|---|
| X | 3.48 | 0.626 | -0.33 | 0.059 | -0.07i | -0.23 i | -0.44 i | 6.08 | X | 0 | 113.6 | 85.8 | n/a |
| Z | 3.16 | 0.013 | 0.189 | 0.662 | 0.38 i | -0.33 i | 0.08 i | -3.91 | Z | 113.6 | 0 | 27.8 | n/a |
| Y | 2.48 | 0.394 | 0.636 | -0.25 | 0.09 i | -0.02 i | 0.11 i | -1.93 | Y | 85.8 | 27.8 | 0 | n/a |
| D | 1.48 | 0 | 0 | 0 | 0.17 | 0.17 | -0.11 | 0 | D | n/a | n/a | n/a | n/a |

(b) q = 0

| State | E (meV) | $\tilde{p}_x$ | $\tilde{p}_y$ | $\tilde{p}_z$ | $\tilde{m}_x$ | $\tilde{m}_y$ | $\tilde{m}_z$ | $\langle g_{CD,n}\rangle$ x10$^5$ | | Y | Z | X | D |
|---|---|---|---|---|---|---|---|---|---|---|---|---|---|
| Y | 3.41 | 0.50 | -0.50 | 0 | -0.15i | -0.15 i | -0.46 i | 0 | Y | 0 | 90.0 | 90.0 | 0.0 |
| Z | 3.18 | 0.136 | 0.136 | 0.657 | 0.37 i | -0.37 i | 0 | 0 | Z | 90.0 | 0 | 0.0 | 90.0 |
| X | 2.53 | 0.53 | 0.53 | -0.27 | 0.05i | -0.05 i | 0 | 0 | X | 90.0 | 0.0 | 0 | 90.0 |
| D | 1.48 | 0 | 0 | 0 | 0.17 | 0.17 | -0.11 | 0 | D | 0.0 | 90.0 | 90.0 | n/a |

(c) q = 0.83

| State | E (meV) | $\tilde{p}_x$ | $\tilde{p}_y$ | $\tilde{p}_z$ | $\tilde{m}_x$ | $\tilde{m}_y$ | $\tilde{m}_z$ | $\langle g_{CD,n}\rangle$ x10$^5$ | | Y | Z | X | D |
|---|---|---|---|---|---|---|---|---|---|---|---|---|---|
| Y | 3.48 | 0.327 | -0.63 | -0.06 | -0.23i | -0.07 i | -0.44 i | -6.08 | Y | 0 | 66.4 | 94.2 | n/a |
| Z | 3.16 | 0.189 | 0.013 | 0.662 | 0.33 i | -0.38 i | -0.08 i | 3.91 | Z | 66.4 | 0 | 27.8 | n/a |
| X | 2.48 | 0.636 | 0.394 | -0.25 | 0.02 i | -0.09 i | -0.11 i | 1.93 | X | 94.2 | 27.8 | 0 | n/a |
| D | 1.48 | 0 | 0 | 0 | 0.17 | 0.17 | -0.11 | 0 | D | n/a | n/a | n/a | n/a |

**Extended Table- 9**. Exciton fine structure and transition dipoles, NCs with orthorhombic facets.

The table shows the energies and cartesian components of the electric and magnetic dipole transition vectors for (a) ferroelectric and (b) centrosymmetric CsPbBr$_3$ NCs bounded by low index orthorhombic lattices planes $\{100\}_o$, $\{010\}_o$, and $\{001\}_o$. The effective edge length is $L_e$=10nm and basal edge length ratio $L_y/L_x = 1.2$. Calculations use the parameters of **Extended Table- 7** but with the polar distortion set to zero for panel (b), while maintaining the same bandgap and exciton parameters as the ferroelectric phase. Coordinates are defined in reference to the NC edges, see schematic in **Fig. 6** panel (e). The set of states are labelled according the major component of the exciton transition dipole. Components of the dimensionless transition dipoles $\tilde{p}_{X_i}$ and $\tilde{m}_{X_i}$ are given, where the full transition dipoles are given by $P_{X_i} = i|P_K|\mathcal{K}\,\tilde{p}_{X_i}$, and $M_{X_i} = \hbar\mathcal{K}\,\tilde{m}_{X_i}$ in terms of $P_K$, the Kane momentum matrix element and $\mathcal{K}$, the overlap factor (Supplementary Eq. S6.5), common to all fine structure levels. The inner product $\tilde{p}\cdot\tilde{m}$, proportional to the rotary strength of an orientationally averaged colloid, is also given for each transition. The angles between the projections into the $x,y$ plane of the electric dipole transition vectors are given for all pairs of states for which electric dipole transition matrix elements are non-zero.

**(a) ferroelectric**

| State | E (meV) | Electric trans. dipoles | | | Magnetic trans. dipoles | | | Im[$\tilde{p}\cdot\tilde{m}$] | | Angles in x,y plane (degrees) | | | |
|---|---|---|---|---|---|---|---|---|---|---|---|---|---|
| | | $\tilde{p}_x$ | $\tilde{p}_y$ | $\tilde{p}_z$ | $\tilde{m}_x$ | $\tilde{m}_y$ | $\tilde{m}_z$ | | | Z | Y | X | D |
| Z | 3.2 | 0.307 | 0 | 0.607 | 0 | -0.52 i | 0 | 0 | Z | 0 | 90 | 0 | n/a |
| Y | 3.19 | 0 | -0.71 | 0 | -0.22i | 0 | -0.46i | 0 | Y | 90 | 0 | 90 | n/a |
| X | 2.77 | 0.708 | 0 | -0.37 | 0 | 0.02 i | 0 | 0 | X | 0 | 90 | 0 | n/a |
| D | 1.48 | 0 | 0 | 0 | 0.24 | 0. | -0.11 | 0 | D | n/a | n/a | n/a | n/a |

**(b) centro-symmetric**

| State | E (meV) | Electric trans. dipoles | | | Magnetic trans. dipoles | | | Im[$\tilde{p}\cdot\tilde{m}$] | | Angles in x,y plane (degrees) | | | |
|---|---|---|---|---|---|---|---|---|---|---|---|---|---|
| | | $\tilde{p}_x$ | $\tilde{p}_y$ | $\tilde{p}_z$ | $\tilde{m}_x$ | $\tilde{m}_y$ | $\tilde{m}_z$ | | | Z | Y | X | D |
| Z | 4.05 | 0.793 | 0 | 0 | 0 | 0 | 0 | 0 | Z | 0 | 90 | 0 | n/a |
| Y | 3.97 | 0 | 0 | 0.814 | 0 | 0 | 0 | 0 | Y | 90 | 0 | 90 | n/a |
| X | 3.94 | 0 | 0.841 | 0 | 0 | 0 | 0 | 0 | X | 0 | 90 | 0 | n/a |
| D | 0 | 0 | 0 | 0 | 0 | 0 | 0 | 0 | D | n/a | n/a | n/a | n/a |

**Extended Table- 10**. Exciton fine structure and transition dipoles: Centrosymmetric CsPbBr$_3$ NCs

The table shows energies and cartesian components of the electric dipole transition vectors for centrosymmetric CsPbBr$_3$ NCs with $L_e$=10nm and different basal edge length ratios $q = L_y/L_x$, calculated using the parameters of **Extended Table- 7**, but with the polar distortion set to zero, while maintaining the same bandgap and exciton parameters as the ferro-electric phase. Coordinates are defined in reference to the NC edges, see schematic in Fig. 6, panel (c) or Supplementary **Error! R eference source not found.**, panel (b). The set of states are labelled according the major component of the exciton transition dipole. Components of the dimensionless transition dipoles $\widetilde{\boldsymbol{p}}_{X_i}$ are given, where the full transition dipole is given by $\boldsymbol{P}_{X_i} = i|P_K|\mathcal{K}\,\widetilde{\boldsymbol{p}}_{X_i}$, in terms of $P_K$, the Kane momentum matrix element and $\mathcal{K}$, the overlap factor (Supplementary Eq. S6.5), common to all fine structure levels. The angles between each of the electric dipoles transition vectors are given for all pairs of states for which electric dipole transition matrix elements are non-zero. All magnetic transition dipoles vanish, consequently so does the orientationally average dissymmetry.

**(a) q = 1.2**

| | Electric transition dipoles | | | | | Angles (degrees) | | |
|---|---|---|---|---|---|---|---|---|
| State | E (meV) | $\widetilde{p}_x$ | $\widetilde{p}_y$ | $\widetilde{p}_z$ | | X | Z | Y |
| X | 4.39 | 0.779 | -0.294 | 0 | X | 0 | 90 | 92.6 |
| Z | 3.97 | 0 | 0 | 0.814 | Z | 90 | 0 | 90 |
| Y | 3.63 | 0.248 | 0.763 | 0 | Y | 92.6 | 90 | 0 |
| D | 0 | 0 | 0 | 0 | | | | |

**(b) q = 0**

| | Electric transition dipoles | | | | | Angles (degrees) | | |
|---|---|---|---|---|---|---|---|---|
| State | E (meV) | $\widetilde{p}_x$ | $\widetilde{p}_y$ | $\widetilde{p}_z$ | | Y | Z | X |
| Y | 4.24 | 0.595 | -0.595 | 0 | Y | 0 | 90 | 90 |
| Z | 3.98 | 0 | 0 | 0.814 | Z | 90 | 0 | 90 |
| X | 3.77 | 0.561 | 0.561 | 0 | X | 90 | 90 | 0 |
| D | 0 | 0 | 0 | 0 | | | | |

**(c) q = 0.83**

| | Electric transition dipoles | | | | | Angles (degrees) | | |
|---|---|---|---|---|---|---|---|---|
| State | E (meV) | $\widetilde{p}_x$ | $\widetilde{p}_y$ | $\widetilde{p}_z$ | | Y | Z | X |
| Y | 4.39 | 0.294 | -0.779 | 0 | Y | 0 | 90 | 87.4 |
| Z | 3.97 | 0 | 0 | 0.814 | Z | 90 | 0 | 90 |
| X | 3.63 | 0.763 | 0.248 | 0 | X | 87.4 | 90 | 0 |
| D | 0 | 0 | 0 | 0 | | | | |